\documentclass{article}

\usepackage{url}            % simple URL typesetting
\usepackage{booktabs}       % professional-quality tables

\usepackage[english]{babel}
\usepackage{natbib}
\usepackage{algorithm}
\usepackage{xcolor}
\usepackage[noend]{algpseudocode}
\usepackage{tikz}
\usepackage{amssymb}
\usepackage[most]{tcolorbox}
\usetikzlibrary{calc,fit,positioning}
\usetikzlibrary{trees, matrix}
\usetikzlibrary{arrows.meta}
\graphicspath{ {./images/} }
\usetikzlibrary{patterns}
\usetikzlibrary{decorations.pathreplacing}
\usepackage{hf-tikz}
\usepackage[margin=1.2in]{geometry}

\tikzset{
    every matrix/.style={
        inner sep=-\pgflinewidth,
        matrix of math nodes,
        column sep=-\pgflinewidth,
        nodes={
            draw=black,
            font=\color{black},
            minimum size=.75cm,
            anchor=center
        }
    }
}

\newcommand{\bX}{\mathbf{X}}

\newcommand{\bx}{\mathbf{x}}
\newcommand{\by}{\mathbf{y}}

\algnewcommand{\Initialization}[1]{%
  \State \textbf{Initialization:}
  \Statex \hspace*{\algorithmicindent}\parbox[t]{.8\linewidth}{\raggedright #1}
}

\title{Considerations in Bayesian agent-based modeling for the analysis of COVID-19 data}

\author{Seungha Um$^\ast$, Samrachana Adhikari\\ [4pt]
\textit{Division of Biostatistics}, 
\textit{Department of Population Health},\\
\textit{New York University Grossman School of Medicine}
\\[2pt]
}

\begin{document}

\maketitle

\begin{abstract}
{Agent-based model (ABM) has been widely used to study infectious disease transmission by simulating behaviors and interactions of autonomous individuals called agents. In the ABM, agent states, for example infected or susceptible, are assigned according to a set of simple rules, and a complex dynamics of disease transmission is described by the collective states of agents over time. Despite the flexibility in real-world modeling, ABMs have received less attention by statisticians because of the intractable likelihood functions which lead to difficulty in estimating parameters and quantifying uncertainty around model outputs. To overcome this limitation, we propose to treat the entire system as a Hidden Markov Model and develop the ABM for infectious disease transmission within the Bayesian framework. The hidden states in the model are represented by individual agent's states over time. We estimate the hidden states and the parameters associated with the model by applying particle Markov Chain Monte Carlo algorithm. Performance of the approach for parameter recovery and prediction along with sensitivity to prior assumptions are evaluated under various simulation conditions. %, in order to identify potential concerns with the ABM approach and to show how to improve it using knowledge‐based prior distributions. 
Finally, we apply the proposed approach to the study of COVID-19 outbreak on Diamond Princess cruise ship and examine the differences in transmission by key demographic characteristics, while considering different network structures and the limitations of COVID-19 testing in the cruise.}
{Agent-based model; Particle filter; Bayesian inference; Hidden Markov Model; Compartmental model.}
\end{abstract}

%we are unable to estimate parameters required for model specification and to quantify uncertainty regarding the model forecasts due to intractable likelihood functions. We use a Bayesian framework to fit the ABM in infectious disease transmission, implementing particle Markov chain Monte Carlo to sample from posterior distribution of parameters given the observed data

\section{Introduction}\label{sec:intro}

Agent-based model (ABM) is a simulation based modeling technique that aims to describe complex dynamic processes, such as the spread of infectious disease. ABM provides considerable flexibility by explaining the complex dynamic process using simple rules that incorporate characteristics of individual entities, called agents, and their interactions. Thus, mechanisms which are often difficult or impossible to model directly at the population level can be incorporated at the smaller scale into the development of ABM \citep{Hooten2010,Grimm2013}. ABM can capture emergent phenomena resulting from the interactions of agents, and can be used to simulate counterfactual outcomes in hypothetical experiments which are impossible or unethical to conduct in the real world. Due to these benefits, ABM has been successfully applied in many fields, including epidemiology \citep{Perez2009,Hooten2020}, sociology \citep{Romano2009,Tom1996}, business \citep{Rand2011,Crowder2012}, and social sciences \citep{YANG2022,Aschwanden2012}.

%Agent-based model(ABM) is a simulation modeling technique that aids to describe individual-based dynamics within a system. 
%In the ABM, a system is modeled as a collection of autonomous objects called agents, and each agent can be described by heterogeneous rules and their interaction associated with the agent's state, the states of other agents and the environment. 
%By simulating the actions of agents over time, we can understand the behavior of the system as well as what determines its outcome. 
%In comparison to other modeling techniques, ABM offers the following advantages: (i) ABM provides considerable flexibility since complex systems are explained by simple rules that incorporate individual characteristic; (ii) ABM incorporates mechanisms at the small scale, which are difficult or impossible to model directly at the population level \citep{Hooten2010,Grimm2013}; (iii) ABM captures emergent phenomena which result from the interactions of individual entities; (iv) ABM is not restricted to observed data, but can also be used to simulate counterfactual or illegitimate experiments. Due to the benefits of ABM, the application of ABMs has been successful in many fields, including epidemiology \citep{Perez2009,Hooten2020}, sociology \citep{Romano2009,Tom1996}, business \citep{Rand2011,Crowder2012} and social sciences \citep{YANG2022,Aschwanden2012}.

Despite the flexibility in real-world modelling, ABMs have received less attention in statistical literature because simulation-based models, such as ABMs, often lack tractable likelihood functions \citep{David2021}. Due to the intractable likelihood functions, %it is difficult to estimate the parameters needed to define the behaviors of an agent and to quantify the uncertainty associated with model predictions, even when longitudinal outcome data, for example infected or susceptible state of an agent over time, may be observed at individual level. The estimation of parameters and uncertainty becomes even more difficult when individual data are not available and may be available at only aggregated population level. 
it is difficult to estimate the parameters needed to define the behaviors of an agent and to quantify the uncertainty associated with model predictions. Consequently, the parameters in ABM are typically derived from other studies. Alternatively, the parameter estimation is available by using methods to approximate intractable likelihood. As public health agencies typically report aggregate data at population level rather than at individual level to protect individuals' privacy, the approximation of likelihood is implemented based on aggregate data. Consider an example of COVID-19 outbreak on Diamond Princess cruise ship \citep{wiki,Moriarty2020}. Although passengers became infected through close contact and developed symptoms at different transmission rates, only cumulative confirmed cases were reported on a daily basis rather than individual infection status over time.

Few studies have investigated approximation of intractable likelihood in the context of ABM. First, the approximated Bayesian computation (ABC) can be considered, where the parameters are estimated by running several simulations for each grid in the parameter space, and then the discrepancies between simulated data and observed data are measured with appropriate metric and tolerance level \citep{Heard2014,Heard2014(2)}. ABC method, however, can be problematic because of the difficulty in determining the metric, tolerance, and summary statistics to decide whether the simulated data is sufficiently close to the observed data. Alternatively, an emulator, which depends on the experimental results from ABM, has been used in place of the ABM and parameters are tuned in order to calibrate to the ABM \citep{David2021}. The use of emulators was examined for the spread of infectious disease such as H1N1, HIV, and COVID-19 by \cite{Farah2014,Hooten2020} and \cite{Heard2014(2)}. Despite the computational efficiency of the emulator compared to ABC approach, emulators can introduce additional uncertainty by adopting surrogate models and are commonly used only for continuous-valued responses\citep{Hooten2020}. 

Alternatively, the Hidden Markov Model (HMM) for the ABM can be developed to estimate parameters and uncertainty around predicted outcomes using Bayesian modeling approaches. In the HMM representation of an ABM, individual agent states, for example presence or absence of infection, are assumed to be hidden (unobserved) states that evolve over time. In many applications of ABM, while the underlying dynamics are individual based, observed data consist of aggregate output, such as the sum of total infected agents. As a result, the HMM represents our best understanding of ABM dynamics where the individual agent's states are treated as hidden and estimated using filtering approaches such as particle filter (PF). The approximated hidden states from PF will then allow us to estimate parameters in the model using Bayesian or maximum likelihood methods. The HMM approach aims to compute the posterior distributions of the agent states in a Markov process and then calculate the likelihood.

In recent years, applications of PF and its extensions have been investigated in the context of ABMs for disease transmission. For example, \cite{kurt2015} applied the PF to ABM for neighborhood-based infection for examining the effect of neighborhood types on the number of infections. This application, however, did not consider parameter estimation and only focused on updating the hidden states assuming known parameters. Alternatively, \cite{ju2021sequential} proposed a modified PF for ABM to estimate parameters associated with the behavior of the agents. In the proposed modified PF, all future observations are considered when proposing possible agent states, as opposed to just the current observation considered in previous versions of the PF. While, consideration of all future observations is helpful for the estimation of likelihood, it can also be computationally challenging, especially when dealing with a large number of agents in real-world data analysis. Further, while non-identifiability and multi-modality of parameters are common issues seen in HMM approach for ABM \citep{Hooten2020}, these issues have not been investigated in the applications of PF and its extensions to HMM for ABM.

In this paper, we conduct an extensive simulation study to explore operating characteristics of the HMM approach for ABM, including parameter estimation and sensitivity to assumptions on prior distribution. To alleviate the computational burden of inference in PF for ABM, we use bootstrap particle filter (BPF) which has efficient computation cost for particle sampling and focus on improving parameter estimation. We use the combination of PF and Markov chain Monte Carlo (MCMC) to estimate parameters given the observed data within Bayesian framework. Moreover, we apply HMM approach to the study of COVID-19 outbreak on Diamond Princess cruise ship and examine the age-dependent effects on transmission considering different network structures and the limitations of COVID-19 testing. Code for implementing our approach is available on \texttt{GitHub} at \url{https://github.com/Seungha-Um/agentSIS}

The rest of the paper is organized as follows. After briefly reviewing the ABM for disease transmission modelling in section \ref{sec::SIS}, we introduce HMM approach for ABM along with details on PF in section \ref{sec:BI}. Section \ref{sec:simulation} includes results of our approach on simulated data under various simulation setting. In Section \ref{sec:real}, we demonstrate the application to the COVID-19 data on Diamond Princess cruise. Section 6 concludes with a discussion. %Details appear in the Appendix and the code for implementing our proposed model is available on \texttt{GitHub} at {\color{blue}\url{https://github.com/Seungha-Um/agentSIS}}

%\subsection{Agent-based model}

\section{Statistical Agent-based SIS model}\label{sec::SIS}
In this section, we review a compartmental model for disease transmission and introduce ABM to allow interaction between the agents and unique disease transition rates for each agent. Further, we describe HMM approach for agent-based compartment model.

\subsection{Compartmental SIS model}
Susceptible-infected-susceptible (SIS) model is a classical framework for modeling the spread of infectious diseases which do not confer any long-lasting immunity. The SIS model divides a population into two states or compartments: susceptible (S) and infected (I). %Agents do not have immunity upon recovery from infection and become susceptible again.
In the SIS model, the disease is transmitted from an infected agent to a susceptible agent with probability proportional to an infected rate $\lambda$ which is interpreted as the average number of contacts per person per time. Also, an infected subject recovers with probability proportional to a recovery rate $\gamma$ where $1/\gamma$ is interpreted as the typical time from infection to recovery. Each subject is considered to be in one compartment at a given time. The classical SIS model to represent the dynamics of total number of susceptible and infected individuals over time can be expressed as
\begin{equation}\label{eq:SIS}
  \begin{split}
    S_{t+1} = S_{t} - \lambda \frac{S_t}{N}  I_t + \gamma I_{t}\\
    I_{t+1} = I_{t} + \lambda \frac{S_t}{N}  I_t -\gamma I_{t}
  \end{split}
\end{equation}
where $S_{t}$ and $I_t$ are the number of susceptible and infectious individuals for discrete time points $t=1,\cdots,T$, respectively. A closed population of size $N$ is considered such that $N = S_t + I_t$ for all $t$. The reproduction number, $R_0 = \lambda/\gamma$, which is interpreted as the expected number of secondary infections from a single infection in a population where all subjects are susceptible, is an important indicator used to assess whether an epidemic is growing, shrinking or remaining stable and to examine the effectiveness of interventions.

The SIS model is a top-down approach where behavioral characteristics of populations are characterized by aggregates of individuals. Although it enables the study of large-scale processes involving homogeneous individuals, the assumptions are too strong; populations are completely mixed and all agents have equal infection and recovery rates. To capture the unique rates of each agent allowing interaction between the agents, the ABM can be considered for modeling disease transmission from a bottom-up approach.

\begin{figure}[t]
\centering
\begin{tikzpicture}
\filldraw [draw=none,fill=gray!20] (-0.5,-0.5) rectangle (3.5,-9.5);
%\filldraw [draw,thick,dotted,fill=gray!20] (-0.5,-0.5) rectangle (3.5,-9.5);

% Input Layer
\foreach \i/\letter in {1/N\i, 2/N\i, 3/N\i}
{\node[rectangle,  
        scale = 0.8, draw,fill=purple!20] (Input-\i) at (0,-\i) {$\letter$};}
\node[rectangle, scale = 0.8, draw,fill=purple!20] at (1.5, -1)   (H1) {N4};
\node[rectangle, draw, minimum size = 6mm, fill=purple!50] at (1.5, -2)(H2) {$X_1^1$};
\node[rectangle, scale = 0.8, draw,fill=purple!20] at (1.5, -3)   (H3) {N5};
\foreach \i /\letter in {1/N6, 2/N7, 3/N8}
{\node[rectangle, 
        scale = 0.8, draw, fill=purple!20] (Output-\i) at (3,-\i) {$\letter$};}
\foreach \source in {1,2,3} \path (Input-\source) edge (H2);
\foreach \dest in {1,2,3} \path (H2) edge (Output-\dest);
\path (H1) edge (H2); \path (H3) edge (H2);
%%%%%%%%%%%%%%%%%%%%%   1-2 
% Input Layer
\foreach \i/\letter in {4/N1, 5/N2, 6/N3}
{\node[rectangle,  
        scale = 0.8, draw,fill=teal!20] (Input-\i) at (0,-\i) {$\letter$};}
 \node[rectangle, scale = 0.8, draw,fill=teal!20] at (1.5, -4)   (H4) {N4};
\node[rectangle, draw,minimum size = 6mm,fill=teal!50] at (1.5, -5)(H5) {$X_1^2$};
\node[rectangle, scale = 0.8, draw,fill=teal!20] at (1.5, -6)   (H6) {N5};
\foreach \i /\letter in {4/N6, 5/N7, 6/N8}
{\node[rectangle, scale = 0.8, draw, fill=teal!20] (Output-\i) at (3,-\i) {$\letter$};}
\foreach \source in {4,5,6} \path (Input-\source) edge (H5);
\foreach \dest in {4,5,6} \path (H5) edge (Output-\dest); \path (H4) edge (H5); \path (H5) edge (H6);
%%%%%%%%%%%%%%%%%%%%%%% 1-3
\foreach \i/\letter in {7/N1, 8/N2, 9/N3}
{\node[rectangle, scale = 0.8, draw,fill=orange!10] (Input-\i) at (0,-\i) {$\letter$};}
\node[rectangle, scale = 0.8, draw,fill=orange!10] at (1.5, -7) (H7) {N4};
\node[rectangle, draw,minimum size = 6mm, fill=orange!40] at (1.5, -8)(H8) {$X_1^3$};
\node[rectangle,scale = 0.8, draw,fill=orange!10] at (1.5, -9) (H9) {N5};
\foreach \i /\letter in {7/N6, 8/N7, 9/N8}
{\node[rectangle, scale = 0.8, draw, fill=orange!10] (Output-\i) at (3,-\i) {$\letter$};}
\foreach \source in {7,8,9}\path (Input-\source) edge (H8);
\foreach \dest in {7,8,9}\path (H8) edge (Output-\dest);
\path (H7) edge (H8);\path (H8) edge (H9);
%%%%%%%%%%%%%%% 2nd column
\filldraw [draw=none,fill=gray!20] (4.5,-0.5) rectangle (8.5,-9.5);
\foreach \i/\letter in {1/N\i, 2/N\i, 3/N\i}
{\node[rectangle, scale = 0.8, draw,fill=purple!20] (Input-\i) at (5,-\i) {$\letter$};}
 \node[rectangle, scale = 0.8, draw,fill=purple!20] at (6.5, -1)   (H21) {N4};
\node[rectangle, draw,minimum size = 6mm, fill=purple!50] at (6.5, -2)(H22) {$X_2^1$};
\node[rectangle, scale = 0.8, draw,fill=purple!20] at (6.5, -3)   (H23) {N5};
\foreach \i /\letter in {1/N6, 2/N7, 3/N8}
{\node[rectangle, scale = 0.8, draw, fill=purple!20] (Output-\i) at (8,-\i) {$\letter$};}
\foreach \source in {1,2,3} \path (Input-\source) edge (H22);
\foreach \dest in {1,2,3} \path (H22) edge (Output-\dest);
\path (H21) edge (H22); \path (H23) edge (H22);
%%%%%%%%%%%%%%%%%%%%%   2-2 
% Input Layer
\foreach \i/\letter in {4/N1, 5/N2, 6/N3}
{\node[rectangle, scale = 0.8, draw,fill=teal!20] (Input-\i) at (5,-\i) {$\letter$};}
\node[rectangle, scale = 0.8, draw,fill=teal!20] at (6.5, -4)   (H24) {N4};
\node[rectangle, draw, minimum size = 6mm, fill=teal!50] at (6.5, -5)(H25) {$X_2^2$};
\node[rectangle, scale = 0.8, draw,fill=teal!20] at (6.5, -6)   (H26) {N5};
\foreach \i /\letter in {4/N6, 5/N7, 6/N8}
{\node[rectangle, scale = 0.8, draw, fill=teal!20] (Output-\i) at (8,-\i) {$\letter$};}
\foreach \source in {4,5,6} \path (Input-\source) edge (H25);
\foreach \dest in {4,5,6}\path (H25) edge (Output-\dest);
\path (H24) edge (H25); \path (H25) edge (H26);
%%%%%%%%%%%%%%%%%%%%%%% 2-3
\foreach \i/\letter in {7/N1, 8/N2, 9/N3}
{\node[rectangle, scale = 0.8, draw,fill=orange!10] (Input-\i) at (5,-\i) {$\letter$};}
 \node[rectangle, scale = 0.8, draw,fill=orange!10] at (6.5, -7) (H27) {N4};
\node[rectangle, draw,minimum size = 6mm, fill=orange!40] at (6.5, -8)(H28) {$X_2^3$};
\node[rectangle, scale = 0.8, draw,fill=orange!10] at (6.5, -9) (H29) {N5};
\foreach \i /\letter in {7/N6, 8/N7, 9/N8}
{\node[rectangle, scale = 0.8, draw, fill=orange!10] (Output-\i) at (8,-\i) {$\letter$};}
\foreach \source in {7,8,9} \path (Input-\source) edge (H28);
\foreach \dest in {7,8,9} \path (H28) edge (Output-\dest);
\path (H27) edge (H28); \path (H28) edge (H29);
%%%%%%%%%%%%%%% 3rd column
\filldraw [draw=none,fill=gray!20] (9.5,-0.5) rectangle (13.5,-9.5);
\foreach \i/\letter in {1/N\i, 2/N\i, 3/N\i}
{\node[rectangle, scale = 0.8, draw,fill=purple!20] (Input-\i) at (10,-\i) {$\letter$};}
\node[rectangle,  scale = 0.8, draw,fill=purple!20] at (11.5, -1)   (H31) {N4};
\node[rectangle, draw, minimum size = 6mm, fill=purple!50] at (11.5, -2)(H32) {$X_3^1$};
\node[rectangle, scale = 0.8, draw,fill=purple!20] at (11.5, -3)   (H33) {N5};
\foreach \i /\letter in {1/N6, 2/N7, 3/N8}
{\node[rectangle,  scale = 0.8, draw, fill=purple!20] (Output-\i) at (13,-\i) {$\letter$};}
\foreach \source in {1,2,3} \path (Input-\source) edge (H32);
\foreach \dest in {1,2,3} \path (H32) edge (Output-\dest);
\path (H31) edge (H32); \path (H33) edge (H32);
%%%%%%%%%%%%%%%%%%%%%   3-2 
\foreach \i/\letter in {4/N1, 5/N2, 6/N3}
{\node[rectangle,  scale = 0.8, draw,fill=teal!20] (Input-\i) at (10,-\i) {$\letter$};}
\node[rectangle, scale = 0.8, draw,fill=teal!20] at (11.5, -4)   (H34) {N4};
\node[rectangle, draw, minimum size = 6mm, fill=teal!50] at (11.5, -5)(H35) {$X_3^2$};
\node[rectangle, scale = 0.8, draw,fill=teal!20] at (11.5, -6)   (H36) {N5};
\foreach \i /\letter in {4/N6, 5/N7, 6/N8}
{\node[rectangle, scale = 0.8, draw, fill=teal!20] (Output-\i) at (13,-\i) {$\letter$};}
\foreach \source in {4,5,6} \path (Input-\source) edge (H35);
\foreach \dest in {4,5,6} \path (H35) edge (Output-\dest);
\path (H34) edge (H35); \path (H35) edge (H36);
%%%%%%%%%%%%%%%%%%%%%%% 3-3
\foreach \i/\letter in {7/N1, 8/N2, 9/N3}
{\node[rectangle, scale = 0.8, draw,fill=orange!10] (Input-\i) at (10,-\i) {$\letter$};}
\node[rectangle, scale = 0.8, draw,fill=orange!10] at (11.5, -7) (H37) {N4};
\node[rectangle, draw,minimum size = 6mm, fill=orange!40] at (11.5, -8)(H38) {$X_3^3$};
\node[rectangle, scale = 0.8, draw,fill=orange!10] at (11.5, -9) (H39) {N5};
\foreach \i /\letter in {7/N6, 8/N7, 9/N8}
{\node[rectangle,  scale = 0.8, draw, fill=orange!10] (Output-\i) at (13,-\i) {$\letter$};}
\foreach \source in {7,8,9} \path (Input-\source) edge (H38);
\foreach \dest in {7,8,9} \path (H38) edge (Output-\dest);
\path (H37) edge (H38); \path (H38) edge (H39);

%%%%%%%% transition arrow for X
\draw [->,>=stealth,thick] (3.5,-2) node[above,xshift=0.5cm] {$\lambda^1$,$\gamma^1$}-- (4.5,-2);
\draw [->,>=stealth,thick] (3.5,-5) node[above,xshift=0.5cm] {$\lambda^2$,$\gamma^2$}-- (4.5,-5);
\draw [->,>=stealth,thick] (3.5,-8) node[above,xshift=0.5cm] {$\lambda^3$,$\gamma^3$}-- (4.5,-8);

\draw [->,>=stealth,thick] (8.5,-2) node[above,xshift=0.5cm] {$\lambda^1$,$\gamma^1$} -- (9.5,-2);
\draw [->,>=stealth,thick] (8.5,-5) node[above,xshift=0.5cm] {$\lambda^2$,$\gamma^2$}-- (9.5,-5);
\draw [->,>=stealth,thick] (8.5,-8) node[above,xshift=0.5cm] {$\lambda^3$,$\gamma^3$}-- (9.5,-8);

%%%%%%%%    node & arrow for Y
\filldraw [draw,thick,dotted,fill=white] (-0.5,-10.5) rectangle (13.5,-11.5);

\draw [->,>=stealth,thick] (1.5,-9.5) node[right,yshift=-.5cm] {$\rho$}-- (1.5,-10.5);
\draw [->,>=stealth,thick] (6.5,-9.5) node[right,yshift=-.5cm] {$\rho$}-- (6.5,-10.5);
\draw [->,>=stealth,thick] (11.5,-9.5) node[right,yshift=-.5cm] {$\rho$}-- (11.5,-10.5);
\node[circle, draw, minimum size = 6mm, fill=blue!30] at (1.5, -11)(y1) {$y_1$};
\node[circle, draw, minimum size = 6mm, fill=blue!30] at (6.5, -11)(y2) {$y_2$};
\node[circle, draw, minimum size = 6mm, fill=blue!30] at (11.5, -11)(y3) {$y_3$};
\end{tikzpicture}
\caption{Graphical illustration of agent-based SIS model for $T=3$ and $n=3$. Each agent has 8 neighbors denoted by $\{N_n^m\}_{m\in(1,8)}$. The agent-based dynamics of disease transmission, which are depicted by shaded boxes, are unobserved (hidden). The aggregate data $(y_1,y_2,y_3)$ are observed.}\label{fig:HMM}
\end{figure}
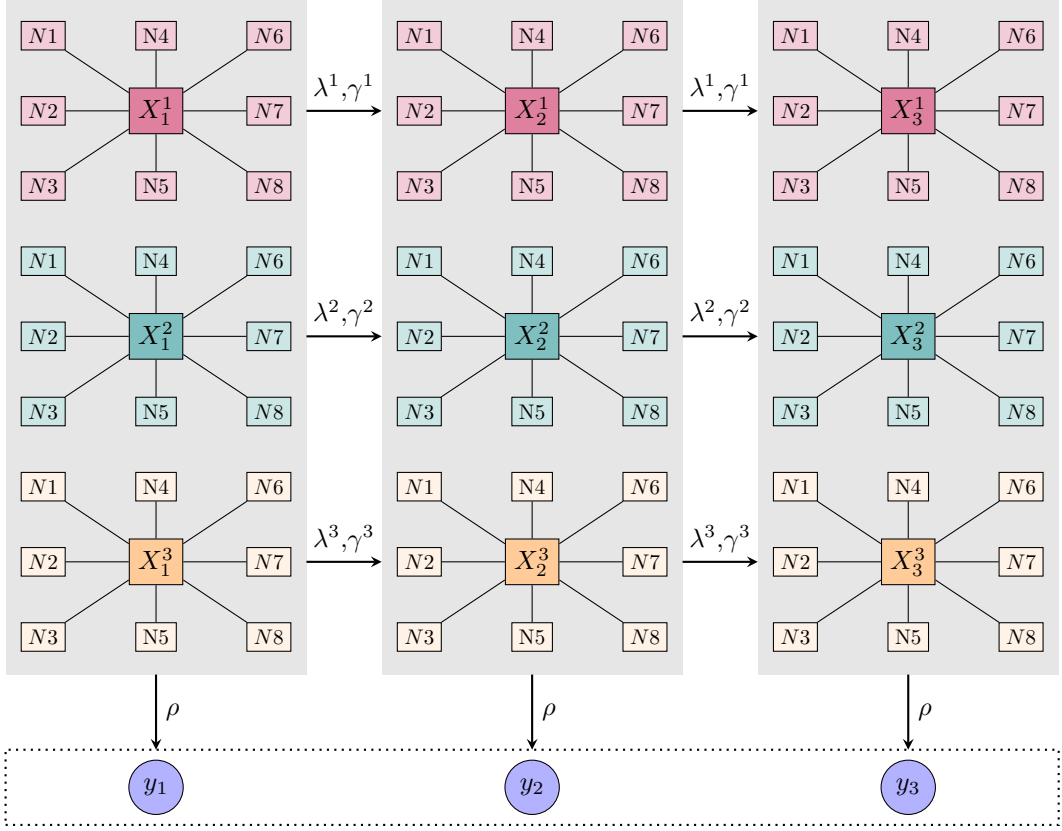

\subsection{Statistical agent-based SIS Model}\label{sec:ABM_SIS}

In the agent-based SIS model, the agent state evolves continuously over time as agents become infected or recover. Let $\mathbf{X}_t = (X_t^1, \ldots, X_t^N)\in\{0,1\}^N$ denote a vector representing states of $N$ agents at time $t$ and $X_t^n$ is the state of an individual agent $n$ at $t$, which takes the value 0 or 1, corresponding to whether the agent is susceptible or infected. Define $\lambda^n$ and $\gamma^n$ as infection rate and recovery rate respectively for $n=1,\cdots,N$, which are unique to each agent and $\mathbf{y}_{0:T} = \{y_0,\cdots,y_T\}$ as the observed (reported) disease prevalence counts at discrete time points $t=1,\cdots,T$. The total number of \emph{actual} infected agents at time $t$ is denoted as $I_t$ which is often unobserved and $\rho$ is the detection probability of the true prevalence.

Figure \ref{fig:HMM} illustrates the graphical representation for agent-based SIS model for three time points with three agents. The agent-based dynamics of disease transmission, which are depicted by shaded boxes, are unobserved (hidden), whereas the aggregate data $(y_1, y_2, y_3)$ are observed.  At $t=1$, the initial agent states $\mathbf{X}_1 = (X_1^1, X_2^1, X_3^1)$ are determined as infected or susceptible according to the corresponding initial infection rate $\alpha_0^n$. Each agent has 8 neighbors denoted by $N_n^m$ for $m=1,\ldots,8$ and the susceptible agent is more likely to become infected when exposed to a large number of infected neighbors.  For time $t=2$ and $t=3$, a susceptible agent $n$ becomes infected with a probability of infection rate $\lambda^n$ adjusted for the number of infected neighbors, whereas the infected agent $n$ becomes susceptible with a probability of recovery rate $\gamma^n$. 

While the underlying dynamics of the disease transmission are governed by agent states $\mathbf{X}_t$, these states are not observed directly. %For example, public health agencies only report aggregate data at the population level rather than at the individual level to protect individuals' privacy; the number of infected agents is reported on a daily basis in COVID-19 data. 
Accordingly, it is suitable to treat the unobserved agent states as hidden states and the observations as reported infection numbers rather than \emph{actual} infection numbers, which connected to the HMM approach. To describe HMM approach for the agent-based SIS model, shown in Figure \ref{fig:HMM}, we adopt a agent-based SIS model proposed by \cite{ju2021sequential} where each individual or group of individuals can have different infection and recovery rates based on individual attributes. The observed prevalence $y_t$ can be modeled as a binomial distribution;
\begin{equation}\label{model:bin}
    y_t \vert I_t,\rho \sim \operatorname{Binomial}(I_t,\rho)
\end{equation}
where the detection probability $\rho$ accounts for the under-reported prevalence; actual case numbers appear vague because of limitations to diagnostic tests, asymptomatic patients and reporting delays in infectious disease such as Covid-19 \citep{Lau2021,Albani2021}.

The observed prevalence at time $t$ depends on unobserved latent agent states $\mathbf{X}_t = (X_t^1, \ldots, X_t^N)$, which can be modeled as an independent Bernoulli distribution;
\begin{equation}\label{model:ber}
    X_t^n \ \vert \ \xi_{t-1}^n \sim \operatorname{Beroulli}(\xi_{t-1}^n).
\end{equation}
where $\xi_{t-1}^n$ corresponds to the transition probability between susceptible and infected state for agent $n$. To allow for the association between disease progression and individual attributes, we define agent-specific initial infection rate $\alpha_0^n$, infection rate $\lambda^n$ and recovery rate $\gamma^n$ as 
\begin{align*}
    \alpha_0^n = \left(1 + \exp(-\beta_{\alpha_0}^Tz^n)\right)^{-1}, \quad
    \lambda^n = \left(1 + \exp(-\beta_\lambda^Tz^n)\right)^{-1}, \quad
    \gamma^n = \left(1+\exp(-\beta_\gamma^Tz^n)\right)^{-1}
\end{align*}
where $\beta_{\alpha_0},\beta_\lambda,\beta_\gamma\in\mathbb{R}^d$ are parameters and $z^n\in\mathbb{R}^d$ are the covariates of agent $n$. 

Given the initial transition probability $\xi_0^n=\alpha_0^n$, the latent state $X_t$ for $t=2,\ldots,T$ evolves according to a Markov process with transition probability \abovedisplayskip14pt\belowdisplayskip14pt
\begin{align*}
    \xi^{n}_{t-1}= \begin{cases}\lambda^{n} \mathcal{D}(n)^{-1} \sum_{m \in \mathcal{N}_n} X_{t-1}^{m}, & \text { if } x_{t-1}^{n}=0 \\ 1-\gamma^{n}, & \text { if } x_{t-1}^{n}=1\end{cases}
\end{align*}
where $\mathcal{N}_n$ is a neighborhood for agent $n$ and $\mathcal{D}(n)$ is the number of neighbors of agent $n$. This transition probability depends only on the last value of the state of agents which is defined as a first-order Markovian. Since attributes of agents $z^n$ account for the infection and recovery rate, the transition probability $\xi_{t-1}^n$ is defined uniquely for each agent. 

\begin{figure}[t]
    \centering
        \includegraphics[width=0.5 \linewidth]{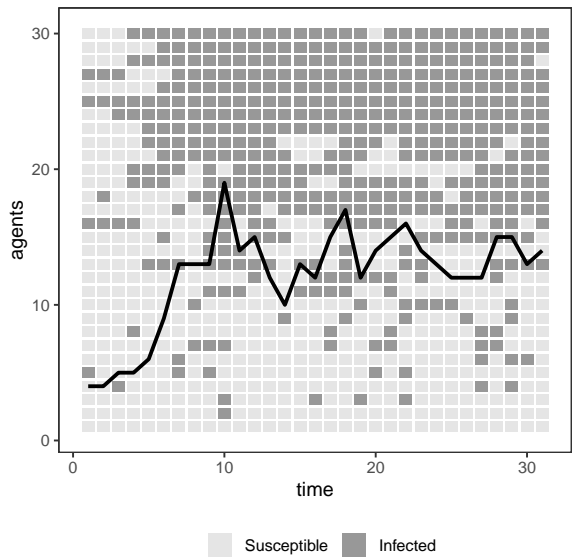}
        \caption{The simulated hidden states and observations from agent-based SIS model. Black line represents observed infection counts and each grid represents agent state which is hidden. The color scale of grid indicates agent state (susceptible or infected) with parameter set of N=30, T=30, $\beta_{\alpha_0}=-log(1/0.2-1)$, $\beta_{\alpha_1}=0$ $\beta_{\lambda_0}=1$, $\beta_{\lambda_1}=2$, $\beta_{\gamma_0}=-1$, $\beta_{\gamma_1}=-3$ and $\rho=0.8$. The y-axis represents agent indices are sorted by magnitude of infection rates.}\label{fig:HMM2}
\end{figure}

Given an initial distribution $p(\mathbf{X}_0|\xi_0)$, a measurement density $p(y_t \vert \mathbf{X}_t=\mathbf{x}_t)$ and a transition density $p(\mathbf{X}_t \vert \mathbf{X}_{t-1} = \mathbf{x}_{t-1},\xi_{t-1})$, the agent-based SIS model can be described as the HMM such that 
\begin{align*}
    &y_t \sim p_\theta(Y_t \vert \mathbf{x}_t,\rho),\\
    &\mathbf{X}_t \sim p_\theta(\mathbf{X}_t \vert \mathbf{x}_{t-1},\xi_{t-1}),\\
    &\mathbf{X}_0 \sim p_\theta(\mathbf{X}_0|\xi_0)
\end{align*}
where $\theta=(\beta_{\alpha_0},\beta_\lambda,\beta_\gamma,\rho)$.

The simulated observations and hidden states from agent-based SIS model is represented in Figure \ref{fig:HMM2} with parameter set of $\beta_{\alpha}=(-log(1/0.2-1), 0)$, $\beta_{\lambda}=(1,2)$, $\beta_{\gamma}=(-1,-3) $ and $\rho=0.8$. The agent covariate $z^n =(z^n_1,z^n_2)$ is assumed, with $z_1^n =1$, and  $z_2^n$ is sampled from a Normal distribution with mean 0 and variance 1. The states of 30 agents evolve according to the Markov process in hidden states and susceptible (infected) agents are indicated by a colored (uncolored) grid. The solid line represents aggregate counts of infection over 30 time points and most of the agents in the upper half are infected over time with higher infection rates. This figure highlights that it is suitable to treat agent states as hidden states and the heterogeneous transition rates between agents cannot be ignored whereas aggregate counts are observed.

%The initial distribution $p(\mathbf{X}_0|\xi_0) = \prod_{n=1}^N \operatorname{Ber}(X_0^n;\xi^n_0)$ can be interpreted as the prior distribution on the initial state of agents given $\xi^n_0$.

\section{Bayesian Inference}\label{sec:BI}
In this section, we illustrate Bayesian inference on the latent variables $\mathbf{X}_{0:T}$ and parameter $\theta$ in the agent-based SIS model. Given the observed prevalence counts $\mathbf{y}_{0:T}$, the complete data likelihood is 
\begin{equation}
p_\theta(\mathbf{x}_{0:T},\mathbf{y}_{0:T}) =p_\theta(\mathbf{y}_{0:T} \vert \mathbf{x}_{0:T}) p_\theta(\mathbf{x}_{0:T}) = \prod_{t=0}^Tp_\theta(y_t\vert\mathbf{x}_t)\prod_{t=1}^T p_\theta(\mathbf{x}_t\vert \mathbf{x}_{t-1})p_\theta(\mathbf{x}_0|\xi_0).
\end{equation}

To estimate the parameters $\theta=(\beta_{\alpha_0},\beta_\lambda,\beta_\gamma,\rho)$, the marginal likelihood $p_\theta(\mathbf{y}_{0:T})$ and sampling of the agent states from a smoothing distribution $p_\theta(\mathbf{x}_{0:T} |\mathbf{y}_{0:T})$ are required. Rather than marginalizing over the latent states to obtain the marginal likelihood $p_\theta(\mathbf{y}_{0:T})$, which is computationally expensive, PF allows us to approximate marginal likelihood and sample hidden states. In the following Sub-sections, we briefly review the PF algorithm and parameter estimation for agent-based SIS model.
 %Thus the posterior inference relies on the posterior distribution $p_\theta(\mathbf{X}_{0:T}\vert y_{0:T})$ and the joint distribution $p(\theta,\mathbf{X}_{0:T} \vert y_{0:T})$. We briefly review the particle filter to approximate posterior distributions of hidden states and how to estimate parameters in the following section. 
 
\subsection{Particle Filter}

%For brevity, we suppress dependence on a parameter set $\theta$ in all following expressions of posterior and conditional posteriors to describe the PF.

The PF algorithm allows to sample latent states $\mathbf{x}_t$ from $p_\theta(\mathbf{x}_{0:t}|\mathbf{y}_{0:t})$ and approximation of the marginal likelihood $p_\theta(\mathbf{y}_{0:t})$ by simulating particles (samples). In the PF algorithm, $p_\theta(\mathbf{x}_{0}|y_{0})$ and $p_\theta(y_0)$ is approximated first and then $p_\theta(\mathbf{x}_{0:1}|\mathbf{y}_{0:1})$ and $p_\theta(\mathbf{y}_{0:1})$ is approximated. The approximation is implemented sequentially until $p_\theta(\mathbf{x}_{0:T}|\mathbf{y}_{0:T})$ and $p_\theta(\mathbf{y}_{0:T})$ are obtained. 

The posterior distribution $p_\theta(\mathbf{x}_{0:t}|\mathbf{y}_{0:t})$ can be approximated by sequential importance sampling when it is computationally intensive or impossible to sample particles directly from the posterior distribution. Using the sequential importance sampling method, $K$ particles can be sampled from a separate distribution called importance distribution, denoted as $q_\theta(x_{0:t}|\mathbf{y}_{0:t})$. This importance distribution includes the support of $p_\theta(\mathbf{x}_{0:t}|\mathbf{y}_{0:t})$. We can then compute importance weight $w\left(\mathbf{x}_{0:t}^{(k)}\right)$ as the ratio $\frac{p_\theta\left(\mathbf{x}^{(k)}_{0:t}|\mathbf{y}_{0:t}\right)}{q_\theta\left(\mathbf{x}_{0:t}^{(k)}|\mathbf{y}_{0:t}\right)}$ where $\mathbf{x}_{0:t}^{(k)}$ represents agent states from each particle $k$. The importance weights can be normalized as $\bar{w}_{t}^{(k)}=w\left(\mathbf{x}_{0: t}^{(k)}\right)/\sum_{j=1}^{K} w\left(\mathbf{x}_{0: t}^{(j)}\right)$ for $k=1,\ldots,K$, and each particle is weighted by corresponding normalized importance weight. Further, using the first-order Markovian property $\bar{w}_{t}^{(k)}$ can be rewritten as 
\begin{align}\label{eq:SIS_weight}
\bar{w}_{t}^{(k)} \propto \bar{w}_{t-1}^{(k)} \frac{p_\theta\left(\mathbf{y}_{t} \mid \mathbf{x}_{t}^{(k)}\right) p_\theta\left(\mathbf{x}_{t}^{(k)} \mid \mathbf{x}_{t-1}^{(k)}\right)}{q_\theta\left(\mathbf{x}_{t}^{(k)} \mid \mathbf{x}_{0: t-1}^{(k)}, \mathbf{y}_{0: t}\right)}.
\end{align}
However, the sequential importance sampling typically fails to represent the posterior distribution due to degeneracy of the importance weights \citep{Doucet2000}. After sampling particles for a few initial time steps, the sequential importance sampling method samples only one particle whose importance weight is close to one while the importance weights of all other particles are close to zero. To avoid the degeneracy of the importance weight, resampling scheme is required \citep{Rubin1987}, which attempts to eliminate particles with small importance weights and to concentrate on particles with large weights. For the resampling, $\{\mathbf{x}^{(k^\prime)}_{0:t}:k^\prime = 1,\cdots,K\}$ is drawn from the set of i.i.d particles $\{\mathbf{x}_{0:t}^{(k)}:k = 1,\cdots,K\}$ with the normalized importance weights in Equation \eqref{eq:SIS_weight}. A common approach for resampling is to use a categorical distribution, denoted by $\mathcal{C}$, that is $\mathcal{C}(\bar{w}_t^{(1)}, \cdots, \bar{w}_t^{(k)})$. 
%Denote the ancestor indexes as $a_{t}^{(p)}$ for $p = 1,\cdots, P$ which consists of indexes of resampled particles at time $t$. 
%The resulting method is PF which is a combination of IS and resampling. 

\begin{figure}[t]
\begin{tikzpicture}[scale=0.7]
\node[text width=3cm] at (2.5, 0.3) (P1) {Particle 1};
\node[text width=3cm] at (7.5, 0.3) (P2) {Particle 2};
\node[text width=3cm] at (12.5, 0.3) (P3) {Particle 3};
\node[text width=3cm] at (17.5, 0.3) (P4) {Particle 4};
\node[text width=3cm] at (1, -3.5) {\small \textit{Weighted by} \\\textit{likelihood}};
\node[text width=3cm] at (1, -7.5) {\small \textit{Resampling}};
\draw[->, thick] (1.5, -2.7) -- (1.5, -4.3) ; 
\draw[->, thick] (6.5, -2.7) -- (6.5, -4.3) ; 
\draw[->, thick] (11.5, -2.7) -- (11.5, -4.3) ; 
\draw[->, thick] (16.5, -2.7) -- (16.5, -4.3) ; 
\draw[->, thick] (1.5, -6.7) -- (1.5, -8.3) ; 
\draw[->, thick] (1.5, -6.7) -- (6.5, -8.2) ; 
\draw[->, thick] (11.5, -6.7) -- (16.5, -8.3) ; 
\draw[->, thick] (16.5, -6.7) --  (11.5, -8.3); 
\filldraw [draw,thick,dotted,fill=gray!2] (-0.7, -0.3) rectangle (3.7, -2.7);
\filldraw [draw,thick,dotted,fill=gray!2] (4.3, -0.3) rectangle (8.7, -2.7);
\filldraw [draw,thick,dotted,fill=gray!2] (9.3, -0.3) rectangle (13.7, -2.7);
\filldraw [draw,thick,dotted,fill=gray!2] (14.3, -0.3) rectangle (18.7, -2.7);
% 1st col
\filldraw [draw,thick,dotted,fill=gray!120] (-0.7, -4.3) rectangle (3.7, -6.7);
\filldraw [draw,thick,dotted,fill=gray!10] (4.3, -4.3) rectangle (8.7, -6.7);
\filldraw [draw,thick,dotted,fill=gray!70] (9.3, -4.3) rectangle (13.7, -6.7);
\filldraw [draw,thick,dotted,fill=gray!60] (14.3, -4.3) rectangle (18.7, -6.7);
% 1st col
\filldraw [draw,thick,dotted,fill=gray!2] (-0.7, -8.3) rectangle (3.7, -10.7);
\filldraw [draw,thick,dotted,fill=gray!2] (4.3, -8.3) rectangle (8.7, -10.7);
\filldraw [draw,thick,dotted,fill=gray!2] (9.3, -8.3) rectangle (13.7, -10.7);
\filldraw [draw,thick,dotted,fill=gray!2] (14.3, -8.3) rectangle (18.7, -10.7);
% 1st col
\node[circle, scale = 0.7, draw, fill = white] at (0, -1) (X1) {$X_t^1$} ;
\node[circle, scale = 0.7, draw, fill = teal!80] at (1, -1) (X2) {$X_t^2$} ;
\node[circle, scale = 0.7, draw, fill = white] at (2, -1) (X3) {$X_t^3$} ;
\node[circle, scale = 0.7, draw, fill = teal!80] at (3, -1) (X4) {$X_t^4$} ;
\node[circle, scale = 0.7, draw, fill = teal!80] at (0, -2) (X1) {$X_t^5$} ;
\node[circle, scale = 0.7, draw, fill = white] at (1, -2) (X2) {$X_t^6$} ;
\node[circle, scale = 0.7, draw, fill = teal!80] at (2, -2) (X3) {$X_t^7$} ;
\node[circle, scale = 0.7, draw, fill = white] at (3, -2) (X4) {$X_t^8$} ;
\node[circle, scale = 0.7, draw, fill = teal!80] at (5, -1) (X5) {$X_t^1$} ;
\node[circle, scale = 0.7, draw, fill = white] at (6, -1) (X6) {$X_t^2$} ;
\node[circle, scale = 0.7, draw, fill = teal!80] at (7, -1) (X7) {$X_t^3$} ;
\node[circle, scale = 0.7, draw, fill = teal!80] at (8, -1) (X8) {$X_t^4$} ;
\node[circle, scale = 0.7, draw, fill = teal!80] at (5, -2) (X5) {$X_t^5$} ;
\node[circle, scale = 0.7, draw, fill = teal!80] at (6, -2) (X6) {$X_t^6$} ;
\node[circle, scale = 0.7, draw, fill = teal!80] at (7, -2) (X7) {$X_t^7$} ;
\node[circle, scale = 0.7, draw, fill = teal!80] at (8, -2) (X8) {$X_t^8$} ;

\node[circle, scale = 0.7, draw, fill = teal!80] at (10, -1) (X10) {$X_t^1$} ;
\node[circle, scale = 0.7, draw, fill = teal!80] at (11, -1) (X21) {$X_t^2$} ;
\node[circle, scale = 0.7, draw, fill = white] at (12, -1) (X22) {$X_t^3$} ;
\node[circle, scale = 0.7, draw, fill = teal!80] at (13, -1) (X9) {$X_t^4$} ;
\node[circle, scale = 0.7, draw, fill = white] at (10, -2) (X10) {$X_t^5$} ;
\node[circle, scale = 0.7, draw, fill = teal!80] at (11, -2) (X21) {$X_t^6$} ;
\node[circle, scale = 0.7, draw, fill = teal!80] at (12, -2) (X22) {$X_t^7$} ;
\node[circle, scale = 0.7, draw, fill = white] at (13, -2) (X9) {$X_t^8$} ;
\node[circle, scale = 0.7, draw, fill = teal!80] at (15, -1) (X26) {$X_t^1$} ;
\node[circle, scale = 0.7, draw, fill = white] at (16, -1) (X27) {$X_t^2$} ;
\node[circle, scale = 0.7, draw, fill = teal!80] at (17, -1) (X23) {$X_t^3$} ;
\node[circle, scale = 0.7, draw, fill = white] at (18, -1) (X25) {$X_t^4$};
\node[circle, scale = 0.7, draw, fill = white] at (15, -2) (X26) {$X_t^5$} ;
\node[circle, scale = 0.7, draw, fill = white] at (16, -2) (X27) {$X_t^6$} ;
\node[circle, scale = 0.7, draw, fill = white] at (17, -2) (X23) {$X_t^7$} ;
\node[circle, scale = 0.7, draw, fill = teal!80] at (18, -2) (X25) {$X_t^8$} ;
% 2nd
\node[circle, scale = 0.7, draw, fill = white] at (0, -5) (X1) {$X_t^1$} ;
\node[circle, scale = 0.7, draw, fill = teal!80] at (1, -5) (X2) {$X_t^2$} ;
\node[circle, scale = 0.7, draw, fill = white] at (2, -5) (X3) {$X_t^3$} ;
\node[circle, scale = 0.7, draw, fill = teal!80] at (3, -5) (X4) {$X_t^4$} ;
\node[circle, scale = 0.7, draw, fill = teal!80] at (0, -6) (X1) {$X_t^5$} ;
\node[circle, scale = 0.7, draw, fill = white] at (1, -6) (X2) {$X_t^6$} ;
\node[circle, scale = 0.7, draw, fill = teal!80] at (2, -6) (X3) {$X_t^7$} ;
\node[circle, scale = 0.7, draw, fill = white] at (3, -6) (X4) {$X_t^8$} ;
\node[circle, scale = 0.7, draw, fill = teal!80] at (5, -5) (X5) {$X_t^1$} ;
\node[circle, scale = 0.7, draw, fill = white] at (6, -5) (X6) {$X_t^2$} ;
\node[circle, scale = 0.7, draw, fill = teal!80] at (7, -5) (X7) {$X_t^3$} ;
\node[circle, scale = 0.7, draw, fill = teal!80] at (8, -5) (X8) {$X_t^4$} ;
\node[circle, scale = 0.7, draw, fill = teal!80] at (5, -6) (X5) {$X_t^5$} ;
\node[circle, scale = 0.7, draw, fill = teal!80] at (6, -6) (X6) {$X_t^6$} ;
\node[circle, scale = 0.7, draw, fill = teal!80] at (7, -6) (X7) {$X_t^7$} ;
\node[circle, scale = 0.7, draw, fill = teal!80] at (8, -6) (X8) {$X_t^8$} ;
\node[circle, scale = 0.7, draw, fill = teal!80] at (10, -5) (X10) {$X_t^1$} ;
\node[circle, scale = 0.7, draw, fill = teal!80] at (11, -5) (X21) {$X_t^2$} ;
\node[circle, scale = 0.7, draw, fill = white] at (12, -5) (X22) {$X_t^3$} ;
\node[circle, scale = 0.7, draw, fill = teal!80] at (13, -5) (X9) {$X_t^4$} ;
\node[circle, scale = 0.7, draw, fill = white] at (10, -6) (X10) {$X_t^5$} ;
\node[circle, scale = 0.7, draw, fill = teal!80] at (11, -6) (X21) {$X_t^6$} ;
\node[circle, scale = 0.7, draw, fill = teal!80] at (12, -6) (X22) {$X_t^7$} ;
\node[circle, scale = 0.7, draw, fill = white] at (13, -6) (X9) {$X_t^8$} ;
\node[circle, scale = 0.7, draw, fill = teal!80] at (15, -5) (X26) {$X_t^1$} ;
\node[circle, scale = 0.7, draw, fill = white] at (16, -5) (X27) {$X_t^2$} ;
\node[circle, scale = 0.7, draw, fill = teal!80] at (17, -5) (X23) {$X_t^3$} ;
\node[circle, scale = 0.7, draw, fill = white] at (18, -5) (X25) {$X_t^4$};
\node[circle, scale = 0.7, draw, fill = white] at (15, -6) (X26) {$X_t^5$} ;
\node[circle, scale = 0.7, draw, fill = white] at (16, -6) (X27) {$X_t^6$} ;
\node[circle, scale = 0.7, draw, fill = white] at (17, -6) (X23) {$X_t^7$} ;
\node[circle, scale = 0.7, draw, fill = teal!80] at (18, -6) (X25) {$X_t^8$} ;
% 3rd
\node[circle, scale = 0.7, draw, fill = white] at (0, -9) (X1) {$X_t^1$} ;
\node[circle, scale = 0.7, draw, fill = teal!80] at (1, -9) (X2) {$X_t^2$} ;
\node[circle, scale = 0.7, draw, fill = white] at (2, -9) (X3) {$X_t^3$} ;
\node[circle, scale = 0.7, draw, fill = teal!80] at (3, -9) (X4) {$X_t^4$} ;
\node[circle, scale = 0.7, draw, fill = teal!80] at (0, -10) (X1) {$X_t^5$} ;
\node[circle, scale = 0.7, draw, fill = white] at (1, -10) (X2) {$X_t^6$} ;
\node[circle, scale = 0.7, draw, fill = teal!80] at (2, -10) (X3) {$X_t^7$} ;
\node[circle, scale = 0.7, draw, fill = white] at (3, -10) (X4) {$X_t^8$} ;
\node[circle, scale = 0.7, draw, fill = white] at (5, -9) (X5) {$X_t^1$} ;
\node[circle, scale = 0.7, draw, fill = teal!80] at (6, -9) (X6) {$X_t^2$} ;
\node[circle, scale = 0.7, draw, fill = white] at (7, -9) (X7) {$X_t^3$} ;
\node[circle, scale = 0.7, draw, fill = teal!80] at (8, -9) (X8) {$X_t^4$} ;
\node[circle, scale = 0.7, draw, fill = teal!80] at (5, -10) (X5) {$X_t^5$} ;
\node[circle, scale = 0.7, draw, fill = white] at (6, -10) (X6) {$X_t^6$} ;
\node[circle, scale = 0.7, draw, fill = teal!80] at (7, -10) (X7) {$X_t^7$} ;
\node[circle, scale = 0.7, draw, fill = white] at (8, -10) (X8) {$X_t^8$} ;
\node[circle, scale = 0.7, draw, fill = teal!80] at (10, -9) (X10) {$X_t^1$} ;
\node[circle, scale = 0.7, draw, fill = white] at (11, -9) (X21) {$X_t^2$} ;
\node[circle, scale = 0.7, draw, fill = teal!80] at (12, -9) (X22) {$X_t^3$} ;
\node[circle, scale = 0.7, draw, fill = white] at (13, -9) (X9) {$X_t^4$} ;
\node[circle, scale = 0.7, draw, fill = white] at (10, -10) (X10) {$X_t^5$} ;
\node[circle, scale = 0.7, draw, fill = white] at (11, -10) (X21) {$X_t^6$} ;
\node[circle, scale = 0.7, draw, fill = white] at (12, -10) (X22) {$X_t^7$} ;
\node[circle, scale = 0.7, draw, fill = teal!80] at (13, -10) (X9) {$X_t^8$} ;
\node[circle, scale = 0.7, draw, fill = teal!80] at (15, -9) (X26) {$X_t^1$} ;
\node[circle, scale = 0.7, draw, fill = teal!80] at (16, -9) (X27) {$X_t^2$} ;
\node[circle, scale = 0.7, draw, fill = white] at (17, -9) (X23) {$X_t^3$} ;
\node[circle, scale = 0.7, draw, fill = teal!80] at (18, -9) (X25) {$X_t^4$};
\node[circle, scale = 0.7, draw, fill = white] at (15, -10) (X26) {$X_t^5$} ;
\node[circle, scale = 0.7, draw, fill = teal!80] at (16, -10) (X27) {$X_t^6$} ;
\node[circle, scale = 0.7, draw, fill = teal!80] at (17, -10) (X23) {$X_t^7$} ;
\node[circle, scale = 0.7, draw, fill = white] at (18, -10) (X25) {$X_t^8$} ;
\end{tikzpicture}
\caption{Illustration of PF strategy for agents SIS model at a fixed time $t$ with 9 agents and 4 particles. Each circle represent agents and infected agents are colored blue. Gray gradient in the second row reflects relative weights of corresponding particles, which is calculated by likelihood assuming $y_t = 4$ and $\rho=0.5$. Particles are re-sampled by the weights.}\label{fig:PF}
\end{figure}
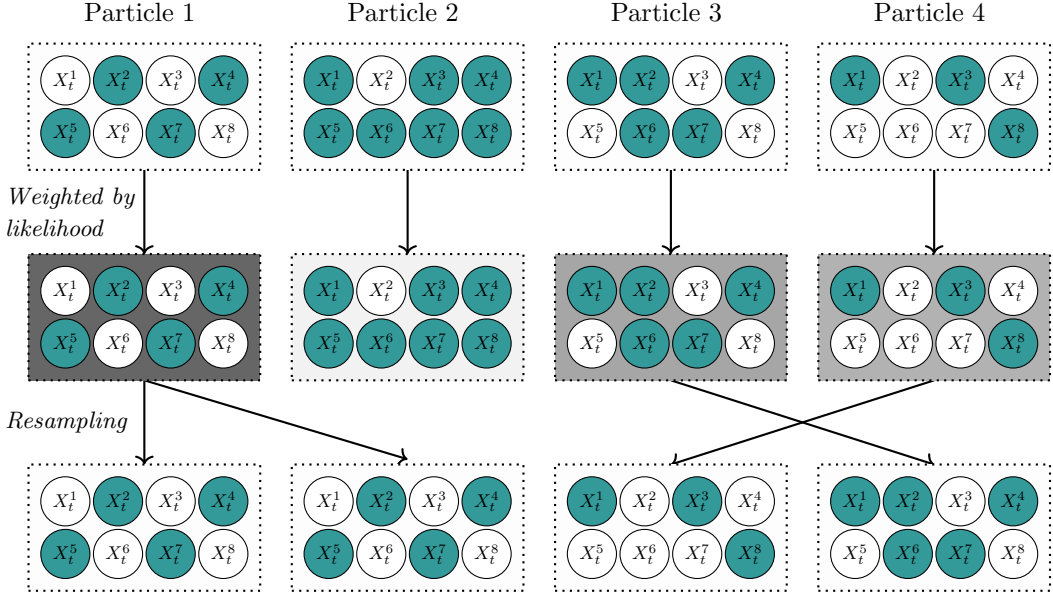

To choose the importance distribution $q_\theta(\mathbf{x}_{0:t}|\mathbf{y}_{0:t})$, the transition prior distribution can be used, which is known as bootstrap particle filter (BPF) \citep{Gordon1993}. Accordingly, the importance weight in Equation \eqref{eq:SIS_weight} is simply computed by likelihood, $\bar{w}_{t}^{(k)} \propto \bar{w}_{t-1}^{(k)} p_\theta\left(y_{t} \mid x_{t}^{(k)}\right)$. The detail of BPF is represented in Algorithm \ref{algo:BPF} and graphical illustration of BPF we developed for the agent-based SIS model is represented in Figure \ref{fig:PF}. In our approach, each particle is composed of a set of agents and it is weighted by the likelihood, which is determined by the number of infected agents within the corresponding particle. Then, the particles are resampled with the weights to avoid the degeneracy of the importance weight. Once particles are sampled until time $T$, the marginal likelihood is estimated by averaging the weights as
\begin{align*}
    \hat{p}(y_t|y_{1:t-1},\theta) = \frac{1}{P} \sum_{p=1}^P w(x_t^{(p)}).
\end{align*}

\begin{algorithm}[t]
\caption{Bootstrap Particle Filter}
%\textbf{Input:} Trajectory $x_{1: T}^{\prime}[r]$ and parameter $\mathbf{\theta} \in \Theta$.\\
%\textbf{Output:} Trajectory $x_{1: T}^{\prime}[r+1]$
\begin{algorithmic}[1] 
    \State Draw $\mathbf{x}_{0}^{(p)} \sim p_{\theta}\left(\mathbf{x}_{0}\vert \xi_0\right)$ for $p=1, \cdots, P$.
    \State Calculate weight $w_{0}\left(\mathbf{x}_{0}^{(p)}\right)=p_{\theta}\left(\mathbf{y}_{0} \mid \mathbf{x}_{0}^{(p)}\right)$ for $p=1, \cdots, P$
    \State Compute normalized weight $\bar{w}_{0}^{(p)}=w_{0}(\mathbf{x}_{0}^{(p)}) / \sum_{j=1}^{P} w_{0}\left(\mathbf{x}_{0}^{(j)}\right)$ for $p = 1,\cdots,P$
    \For{$t=1$ to $T$}
        \State Draw $a_t^{(p)}\sim \mathcal{C}\left(\{\bar{w}^{(p)}_{t-1}\}_{p=1}^{P}\right)$ where $\mathcal{C}$ is the categorical distribution.
        \State Draw $\mathbf{x}_t^{(p)}\sim p\left(\mathbf{x}_t | \mathbf{x}_{t-1}^{a_t^{(p)}},\mathbf{y}_t\right)$ for $p=1,\cdots,P$.
        \State Calculate weight $w_{t}\left(\bx_t^{(p)}\right)=p_{\theta}\left(\mathbf{y}_{t} \mid \mathbf{x}_{t}^{(p)}\right) $ for $p=1, \cdots, P$
        \State Compute normalized weight $\bar{w}_{t}^{p}=w_{t}^{p} / \sum_{j=1}^{P} w_{t}^{j}$ for $p=1, \cdots, P$
\EndFor\textbf{end for}
\end{algorithmic}\label{algo:BPF}
\end{algorithm}

\subsection{Particle Markov Chain Monte Carlo}

Particle Markov Chain Monte Carlo (PMCMC) methods \citep{Andrieu2010} embeds the PF within the Markov chain Monte Carlo (MCMC) algorithm and so are used to carry out inference on hidden states and the parameters in HMM. In PMCMC, the posterior distribution of the hidden states is approximated by PF and the parameters are estimated using the MCMC algorithms. The Metropolis-Hastings (MH) and Gibbs sampler are efficient and widely used  MCMC algorithms within the Bayesian framework, which we adopt for parameter estimation and inference on the hidden states in the proposed agent-based SIS model. 

A method of PMCMC using the MH sampler, called Particle marginal Metropolis-Hastings (PMMH), jointly updates the parameter set $\theta$ and the hidden states $\bX_{0:T}$. In the PMMH algorithm as illustrated in Algorithm \ref{algo:PMMH}, a new parameter set $\theta^{*}$ is sampled from the proposal distribution $g(\cdot\vert\theta)$ and then the hidden states $\bX_{0:T}$ are updated using PF given $\theta^{*}$, which allows the marginal likelihood $\hat{p}\left(\by_{0:T} \mid \theta^{*}\right)$ to be estimated. The proposal $\bX_{0:T}^{*}$, $\theta^{*},$ and $\hat{p}\left(y_{0:T} \mid \theta^{*}\right)$ are accepted or rejected jointly according to a Metropolis-Hastings acceptance ratio. %The PMMH algorithm is illustrated in Algorithm \ref{algo:PMMH}.

\begin{algorithm}[t]
\caption{Particle Marginal Metropolis-Hastings}
\begin{algorithmic}[1] 
    \Initialization{\strut Set $\theta[0]$ arbitrarily \\\vspace{0.4cm}
    Run a particle filter and sample $X_{0:T}[0]\sim\hat{p}(\cdot\vert y_{0:T}, \theta[0])$}\vspace{0.4cm}
    \For{$m=1$ to $M$}
        \State Sample $\theta^*\sim g(\cdot\vert \theta[m-1])$.
        \State Run a particle filter targeting $p_{\theta^*}(x_{0:T}\vert y_{0:T})$ and sample $X_{0:T}^*\sim\hat{p}(\cdot\vert y_{0:T}, \theta^*)$.
        \State With probability
            \begin{align*}
             1\wedge  \frac{\hat{p}(y_{0:T},\theta^*)p(\theta^*)}{\hat{p}(y_{0:T},\theta[m-1])p(\theta[m-1])} \frac{g(\theta[m-1]\vert\theta^*)}{g(\theta^*\vert\theta[m-1])}
            \end{align*}
        set $\theta[m] = \theta^*$, $X_{0:T}[m]=X^*_{0:T}$ and $\hat{p}(y_{0:T},\theta[m]) = \hat{p}(y_{0:T},\theta^*)$; otherwise $\theta[m] = \theta[m-1]$, $X_{0:T}[m]=X_{0:T}[m-1]$ and $\hat{p}(y_{0:T},\theta[m]) = \hat{p}(y_{0:T},\theta[m-1])$\vspace{0.4cm}
\EndFor\textbf{end for}
\end{algorithmic}\label{algo:PMMH}
\end{algorithm}

%targeting $p(\theta,x_{0:T}|y_{0:T})$
Alternatively, a method of PMCMC using the Gibbs sampling, called Particle Gibbs (PG), consists of two steps;
\begin{align*}
    &\text{Draw } \theta^* | \bx_{0:T} \sim p(\theta | \bx_{0:T}, \by_{0:T})\\
    &\text{Draw } \bx_{0:T} | \theta^*\sim p(\bx_{0:T}|\by_{0:T},\theta^*)
\end{align*}
which are performed iteratively. The first step is straightforward with conjugate priors of parameters or can rely on MH algorithm to approximate the posterior distribution $\theta$ \citep{Lindsten2012}. The second step is approximated by PF, but one particle trajectory is set deterministically to a reference trajectory \citep{Andrieu2010}. A reference particle $\bx^\prime_{1: T}$ should be prespecified and should survive throughout all the resampling steps. The reference particle $\bx^\prime_{0:T}$ is sampled from among the particle trajectories with $\mathbb{P}\left(x_{1: T}^\prime=x_{1: T}^{(p)}\right) \propto w_{T}^{(p)}$ from the previous MCMC iteration. The details of PG appear in Appendix (Algorithm \ref{algo:cSMC}). In Section \ref{sec:sim_PG_PMMH}, we compare the performance of PG and PMMH in the agent-based SIS model.

%When there exists a strong dependence between the states and the parameters, the PMMH might performs better since the PG samples  parameters conditioned on the states, and vice versa .\citep{Lindsten2011} In section \ref{sec:simulation}, we show the PMMH outperforms the PG for our model. 
%In agent-based SIS model, conjugate priors are unavailable except for the reporting rate $\rho$ and all parameters are highly correlated with the states.

% An additional ancestor sampling step to PG gives a significant improvement to mixing and alleviates the path degeneracy problem with invariance properties which is known as Particle Gibbs with Ancestor Sampling (PGAS) \citep{Lindsten2014}. At time $t \geq 2$, the PGAS assigns one of the particles $\left\{\pX_{1: t-1}^{i}\right\}_{i=1}^{P}$ to the reference particle $\pX_{t: T}^{\prime}$ creating a history for the partial reference particle. A new value to $a_t^P$ is assigned by the weights
% $$
% \mathbb{P}(a_t^P=j)\propto w_{t-1}^{j} p_{\theta}\left(x_{t}^{\prime} \mid x_{t-1}^{j}\right)
% $$
% for $j=1, \ldots, P$. 
%The parameter set $\Theta$ in the Agent-based SIS model is replaced by $\theta$ in this section for generality.

\section{Simulation Study}\label{sec:simulation}
In this section, we explore operating characteristics of the agent-based SIS model introduced in Section \ref{sec::SIS}. We aim to identify when the HMM approach for ABM can be problematic in terms of parameter estimation through exploratory analysis under various simulation settings and provide how to resolve the problems. 

%In this section, exploratory analysis is provided under various simulation settings. to identify when the HMM approach for ABM has identifiability or multimodality issues. In this section, we explore operating characteristics of the agent-based SIS model introduced in Section \ref{sec::SIS}. By conducting an extensive simulation study, we assess parameter estimation and identification as well as sensitivity to assumptions on prior distributions for both PMMH and PG algorithms discussed above.

In first simulations, the recovery rate is assumed to be known to simplify the model, and so the different settings or performances of PMCMC are compared according to the estimation of transmission rate and reporting rate. As we isolate issues on parameter estimation using non-informative priors and find a suitable PMCMC method for ABM, we extend the case that recovery rates are also unknown. The simulation study is conducted under four broad settings. In the first setting, we examine the influence of number of particles, time points and agent numbers on the performance of PMMH, assuming non-informative priors. In the second simulation setting, we examine the role of informative versus non-informative priors on parameter estimation using PMMH. Finally, in the final two sets of simulation we compare performance of PG and PMMH, and assess parameters estimation when recovery rate is also unknown.

We generate data according to the agent-based SIS model introduced in section \ref{sec:ABM_SIS}. For all simulation settings, we consider $\beta_{\alpha_0}=(-\log(1/0.05-1), 0)$, which corresponds to the initial infection rate of $\alpha_0^n =0.05$ for all $n$, $\beta_{\lambda}=(-1, 2)$, and the reporting rate of $\rho=0.8$. We assume daily time intervals and a fully connected network of agents with degree $n$. A two dimensional agent covariate $z^n=(z_1^n, z^n_2)$ is assumed, with $z_1^n=1$, and $z^n_2$ is sampled from a Normal distribution with mean 0 and variance 1. In simulation 2, we also introduce a binary covariate to reflect that different infection rates for each group are of interest in many real world applications.

All simulation settings are provided with the same initial values for MCMC sampler which is generated from corresponding prior distribution. Our implementation of the PMMH algorithm uses independent Normal random walk proposal distributions for parameters adjusting for acceptance rates between 15\% and 20\%. Parameters are proposed and updated jointly, and 100,000 MCMC samples after a burn-in of 50,000 draws are used for each simulation.

\subsection{Simulation 1 - The number of particles, time points and agents}\label{sec:sim1}
\begin{figure}[t]
    \centering
        \includegraphics[width=1 \linewidth]{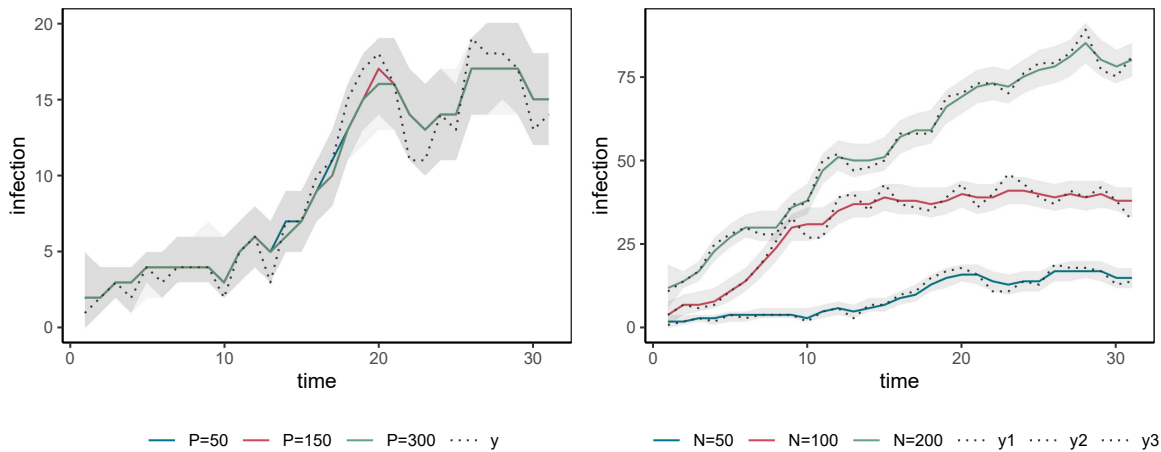}
    \caption{The estimated infection counts varying particle numbers $P\in(50,150,300)$ when $N=50$ and $T=30$ (left) and varying agent numbers $N\in(50,100,200)$ when $T=30$ and $P=100$ (right).}\label{fig:infect_counts}
\end{figure} 
In the first set of simulations, we examine how the number of particles, time points, and number of agents affect the performance of PMCMC for the agent-based SIS model. In this set of simulations, we are concerned with estimation of unknown reporting rate and parameters associated with infection rate. We will also compare the estimated trends in number of infections from the posterior predictive distribution with the underlying true numbers. 

Here, we explore nine different simulations; i) varying number of particles with $P\in(50,150,300)$, with $N=100$ and $T=30$, ii) varying number of time points $T\in(30,100,200)$, with $P=100$ and $N=100$, and iii) varying number of agents $N\in (50,100,200)$, with $T=30$ and $P=100$. We further assume that the average time until recovered is known and fixed as 10 days, i.e. $\lambda^n = 1/10$. Estimation with unknown recovery rate is presented in Sub-section \ref{sec:sim_fully}. The PMMH algorithm is used for sampling parameters and hidden states. We adopt independent Normal prior with standard deviation of 3 and mean of 0 for each component of parameter set $\theta=(\beta_{\alpha_0}, \beta_{\alpha_1},\beta_{\lambda_0}, \beta_{\lambda_1}, \log(\rho/(1-\rho)) )$.

In Figure \ref{fig:infect_counts}, the posterior means of infection counts with varying particle numbers (left) and varying agent numbers (right) for fixed time steps are shown. Regardless of the particle numbers or agent numbers, the 95\% credible interval (CI) of the estimated infection counts includes the true infected counts (represented by gray solid line), and the posterior mean is close to the true counts. The results suggest that we are accurately able to recover the true infection counts with PMMH algorithm and the number of particles and agents do not make a difference in the estimation of posterior mean and CIs of infection counts. 

% the left panel represents the posterior mean of infection counts for varying particle numbers $P\in(50,150,300)$ when $N=50$, $T=30$. Regardless of the particle numbers, the 95\% credible interval (CI) includes the true infected counts (represented by gray solid line) over all time points and the posterior mean is close to the true counts. The number of particles do not make a difference in posterior mean and CIs of infection counts. 

% The right panel in Figure \ref{fig:infect_counts} represents the posterior mean of infection counts over time when $P=100$, $T=30$ varying the number of agents $N\in(50,100,200)$. Likewise, all credible intervals include the true infected counts over all time points and the posterior mean is close to the true counts. The results show that shows we are recovering well the true infection counts with PMMH algorithm. 

\begin{figure}[t]
    \centering
        \includegraphics[width=1 \linewidth]{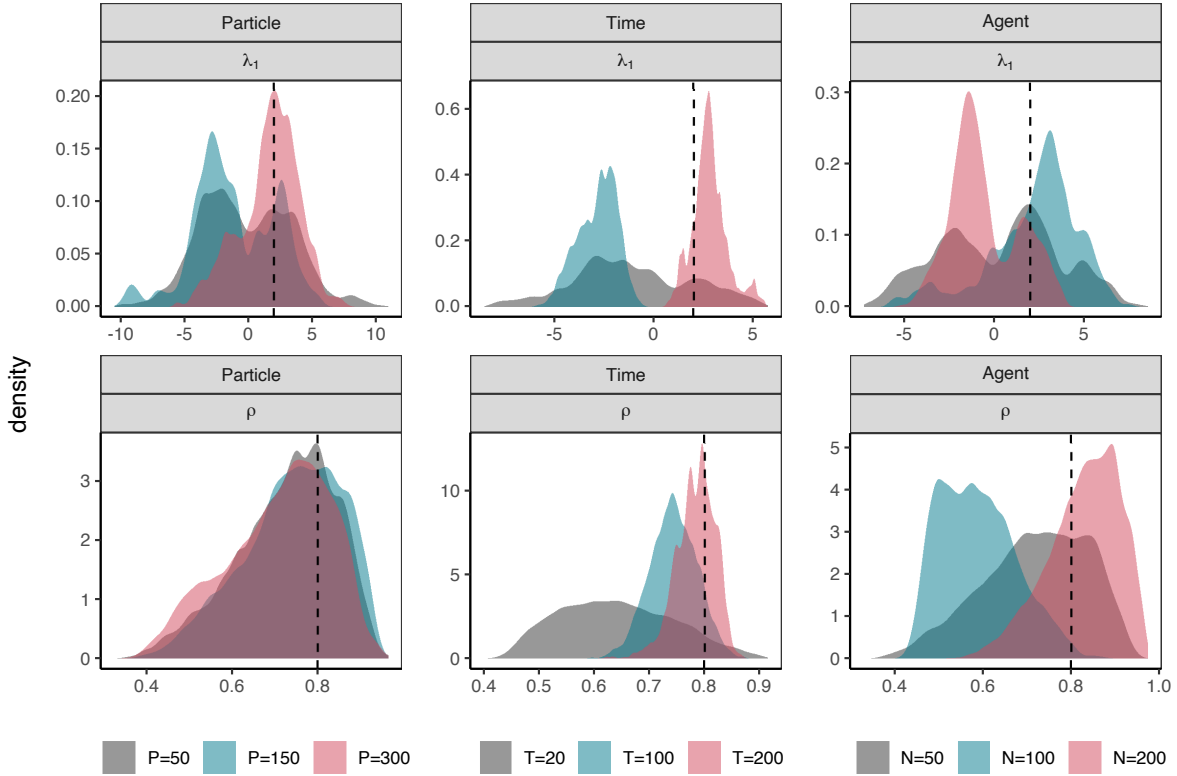}
    \caption{Simulation study 1: The estimated posterior distributions with uninformative priors under 9 different settings. i) the number of particles varies $P\in(50, 150,300)$ at N=50 and T=30 ii) the number of time points varies $T\in(20,100,200)$ at N=100 and P=100 iii) the number of agents varies $N\in(50,100,200)$ at T=30 and P=100.}\label{fig:est_uninfo}
\end{figure}

Next, Figure \ref{fig:est_uninfo} shows the estimated posterior distributions of $(\beta_{\lambda_1},\rho)$ compared with the data generating values from each of the nine corresponding simulations. The estimated posterior distributions of the remaining parameters ($\beta_{\alpha_0}, \beta_{\alpha_1},\beta_{\lambda_0}$) and posterior means of all parameters with 95\% CI are reported in the appendix (Figure \ref{fig:remain_uninfo} and Table \ref{table:comprehensive_Normal}).

The left panel of Figure \ref{fig:est_uninfo} shows that the number of particles does not affect the parameter estimation performance for reporting rate $\rho$. The posterior distributions of $\beta_{\lambda_1}$, however, exhibit bimodality distribution, particularly in the case of small particle numbers. When $P=300$, the posterior distribution of $\beta_{\lambda_1}$ is centered around the true parameter value of 2 but 95\% CI still includes -2. From the middle panel of Figure \ref{fig:est_uninfo}, when $T=20$ and $T=100$, the identifiability issue still remains, showing that the posterior distribution of $\beta_{\lambda_1}$ is centered around $-2$ or includes both $-2$ and $2$. However, provided large number of time points ($T=200$), the posterior mean of parameters are close to the true parameter value without the bimodality issue in $\beta_{\lambda1}$. Also, the result shows that the posterior estimates of $\beta_{\lambda1}$ is closely related to those of estimates of $\rho$ since both posterior distribution for $\beta_{\lambda1}$ and $\rho$ are wide when $T=20$. Consequently, the underlying dynamics are easily identified when the responses are observed over a longer period of time. 

Next, the effect of agent number on parameter estimation is illustrated in the right panel of Figure \ref{fig:est_uninfo}. Regardless of agents numbers, the posterior estimates of $\beta_{\lambda_1}$ represent bimodality issue and the posterior estimates of $\rho$ indicate overestimation or underestimation problem. The result indicates that the increased number of agents cannot resolve the bimodality or identifiability issue in the posterior distribution of $\beta_{\lambda_1}$ under short time period $T = 30$. 

The bimodality issue may have occurred because the distributions of infection rate with $\beta_{\lambda_1}=-2$ and $\beta_{\lambda_1}=2$ are indistinguishable in the hidden state if the individual attribute $z^n_2$ is symmetric at $0$ and uniformly distributed. In this case, the estimation may be sensitive to initial values for the MCMC, and the posterior chain could get stuck in the local optimum. We may require knowledge-based informative priors to establish reasonable ranges for the parameters, which we explore in the next set of simulations.
 %Thus parameter estimation is affected by the number of particle, agents and time points resulting in poor performance. 

% For the large number of agents ($N=200$), the posterior mean of $\beta_{\lambda_1}$ is 2.216 close to true value of 2 but the 95\% CI is too wide (-4.112, 5.9548) including 0. Also, when $N=100$, the the reporting rate $\rho$ is underestimated leading to the overestimated $\beta_{\lambda_1}$. The result indicates that the increased number of agents cannot resolve the bimodal or identifiability issue in the posterior distribution of $\beta_{\lambda_1}$. 

\subsection{Simulation 2 -  knowledge-based prior distribution}\label{sec:prior} 

\begin{figure}[t]
    \centering
        \includegraphics[width=1 \linewidth]{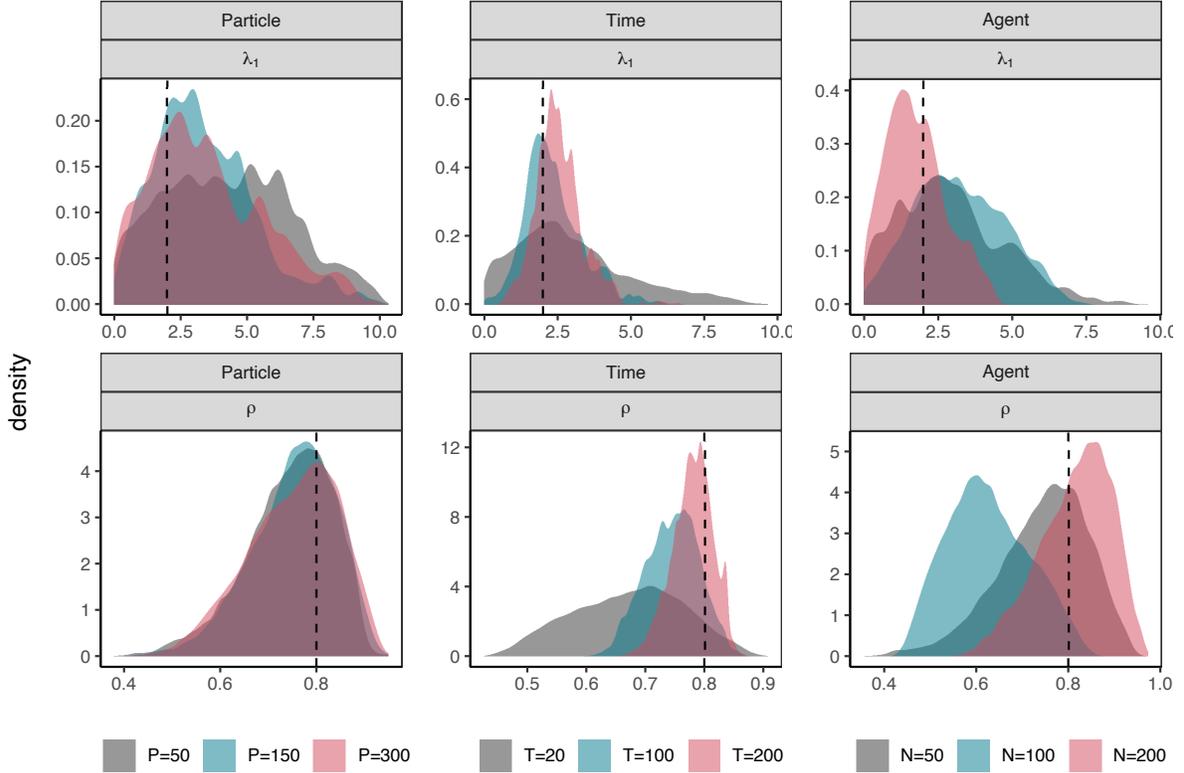}
    \caption{Simulation study 2: The posterior distributions with knowledge-based priors under 9 different settings. i) the number of particles varies $P\in(50, 150,300)$ at N=50 and T=30 ii) the number of time points varies $T\in(20,100,200)$ at N=100 and P=100 iii) the number of agents varies $N\in(50,100,200)$ at T=30 and P=100.}\label{fig:est_knowledge}
\end{figure} 
%When we don't have much information on the parameters, uninformative priors are preferred reflecting our uncertainty. However, in PMMH approach for agent-based SIS model, the uninformative priors lead the particles to spread out too much resulting in poor mixing\cite{Wigren2019} and identifiability or multimodality issues are observed in section \ref{sec:sim1}. To avoid the issues, knowledge-based informative priors can be adopted to restrict range for some parameters \cite{Amanda2016,Lindsten2012}. 
In the second set of simulations, we investigate whether the knowledge-based informative priors can improve the performance of PMMH under the similar setups as in Simulation 1 described in Sub-section \ref{sec:sim1}. %Since it is relatively easy to define whether the relationship between infection rate and individual attribute $z$ is positive or negative, we consider the truncated Normal distribution $N_+(0,3^2)$ for $\beta_{\lambda_1}$ where $N_+(\mu,\sigma^2)$ denotes $N(\mu,\sigma^2)$ truncated with the support of $(0,\infty)$. 
However, more informative prior distributions are considered. We consider the truncated Normal prior distribution $N_+(0,3^2)$ for $\beta_{\lambda_1}$ where $N_+(\mu,\sigma^2)$ denotes $N(\mu,\sigma^2)$ truncated with the support of $(0,\infty)$ in light of the positive correlation between infection rate and individual attribute $z^n$. The truncated Normal distribution $N_-(\mu,\sigma^2)$ can be defined similarly with the support of $(-\infty,0)$. Also, the Normal prior $N(logit(0.8), 1$) is assigned to $logit(\rho)$ which makes $\rho$ fall between 0.360 and 0.965 with probability 0.95. Figure \ref{fig:est_knowledge} shows the estimated posterior distributions of $(\beta_{\lambda_1},\rho)$ compared with the data generating values from the corresponding setting. The estimated posterior distributions of remaining parameters (Figure \ref{fig:remain_knowledge}) and posterior means of all parameters with 95\% CI (Table \ref{table:comprehensive_truncated}) are shown in the appendix.

The left panel of Figure \ref{fig:est_knowledge} shows that the particle numbers do not have a significant impact on the posterior distribution when $N=50$ and $T=30$. The results indicate that, in certain simulation settings, an increase in particle number cannot guarantee improvement in parameter estimation. From the middle panel of Figure \ref{fig:est_knowledge}, the posterior distributions of both $\beta_{\lambda_1}$ and $\rho$ are centered around the true parameter value as the number of time points increases. This result reconfirms that the underlying dynamics are easily identified when the the responses are observed over a long period of time. Next, the right panel of Figure \ref{fig:est_knowledge} shows the effect of agent number on parameter estimation. The large number of agents improves the parameter estimation providing narrow 95\% CIs around the true parameter value, which indicate that the large number of agents can be helpful to identify underlying dynamics over short period time.  

\begin{table}[t]
\centering
{\tabcolsep=2pt
\begin{tabular}{lllllll}
  \toprule
 & & $\beta_{\alpha_0}$ & $\beta_{\alpha_1}$ & $\beta_{\lambda_0}$ & $\beta_{\lambda_1}$ & $\rho$ \\ 
  \midrule
 & True &-2.99 & 0 & -1& 2 & 0.8 \\
    \midrule
   Continuous & mean & -3.83 & 0.34 & -0.13 & 3.23 & 0.63 \\ 
 & 95\% CI &(-7.93,-2.20)  & (-3.34,4.89) & (-1.42,1.39) & (0.52,6.19) & (0.47,0.80) \\ 
   \midrule
% Categorical & mean & -3.173 & -0.499 & -1.081 & 1.875 & 0.850 \\ 
%  & 95\% CI & (-4.610, -2.318) & (-2.370, 1.520) & (-1.517,-0.578) & (0.136, 4.639) & (0.791, 0.912) \\ 
 Categorical & mean & -3.15 & -0.09 & -0.98 & 1.38 & 0.79 \\ 
 & 95\% CI & (-4.28,-2.24) & (-1.91,1.71) & (-1.61,-0.38) & (0.05,4.46) & (0.67,0.91) \\ 
   \bottomrule
\end{tabular}}
\caption{Simulation2: the posterior mean of parameters with 95\% credible interval when $z^n$ is a continuous variable and a categorical variable with $N=100$, $T=30$ and $P=100$.}\label{table:categorical}
\end{table}

Additionally, we examine the parameter estimation when $z^n$ is a binary variable using the same knowledge-based priors. This setting indicates that two groups have different infection rates, which is consistent with our interest in data from the Diamond Princess cruise. The $z^n$ is set as 1 with probability 0.4; otherwise $z^n$ is set as 0. We keep the same parameter values which generate different infection rates such that $\lambda^n = 0.27$ for $z^n=0$ group and $\lambda^n = 0.73$ for $z^n=1$ group. We set $N=100$, $T=30$ and $P=100$ to mimic the Diamond Princess  cruise data, where the confirmed cases are observed daily for 30 days and agent numbers are greater than time points $(N>T)$. The posterior means with 95\% CIs are displayed in Table \ref{table:categorical} compared with those obtained when $z^n$ is a continuous variable with $N=100$, $T=30$ and $P=100$. The estimated infection rates are 0.27 (95\% : 0.17, 0.41) for $z^n=0$ and 0.60 (95\% : 0.33, 0.96) for $z^n=1$. Regardless of the type of $z^n$, the CIs of all parameters contain the true parameter values, but a narrow 95\% CI is provided when $z^n$ is a binary. Our result demonstrates that infection rates by group can be estimated with less posterior uncertainty, provided truncated range for $\beta_{\lambda_1}$ and $\rho$ in spite of short time period.

%when infection rates are defined uniquely for each group. The two different infection rates by group can be defined as a binary variable of $z^n$. We keep the same parameter values which generate different infection rates such that $\lambda^n = 0.27$ for $z^n=0$ and $\lambda^n = 0.73$ for $z^n=1$. 

%The complexity of the model remains unchanged but discrete value of $z^n$ can avoid the identifiability issue between $\beta_{\lambda_0}$ and $\rho$. 

% When $z^n$ is a continuous variable, we can improve parameter estimation in terms of narrow CIs including true values for $\beta_{\alpha_0}$, $\beta_{\alpha_1}$ and $\beta_{\lambda_1}$ despite small number of time points. However, the result shows the overestimated $\beta_{\lambda_0}$ and the underestimated $\rho$. Since $E(y_t) = \rho \sum_{i=1}^N X_t^{i}$, the overestimated intercepts $\beta_{\lambda_0}$ leads to increased infected agents and the underestimated reporting rate $\rho$. We observe that this identifiability issue between $\rho$ and $\beta_{\lambda_0}$ when the number of agents is larger than the number of time points (N $>$T).

% \begin{figure}[t]
%     \centering
%         \includegraphics[width=1 \linewidth]{fig/truncated.eps}
%         \includegraphics[width=1 \linewidth]{fig/cate_est.eps}
%         \caption{truncated normal distribution when $z^n$ is a continuous variable (top) and when $z^n$ is a categorical variable (bottom).}\label{fig:truncated}
% \end{figure} 

\begin{figure}[t]
    \centering
        \includegraphics[width=1 \linewidth]{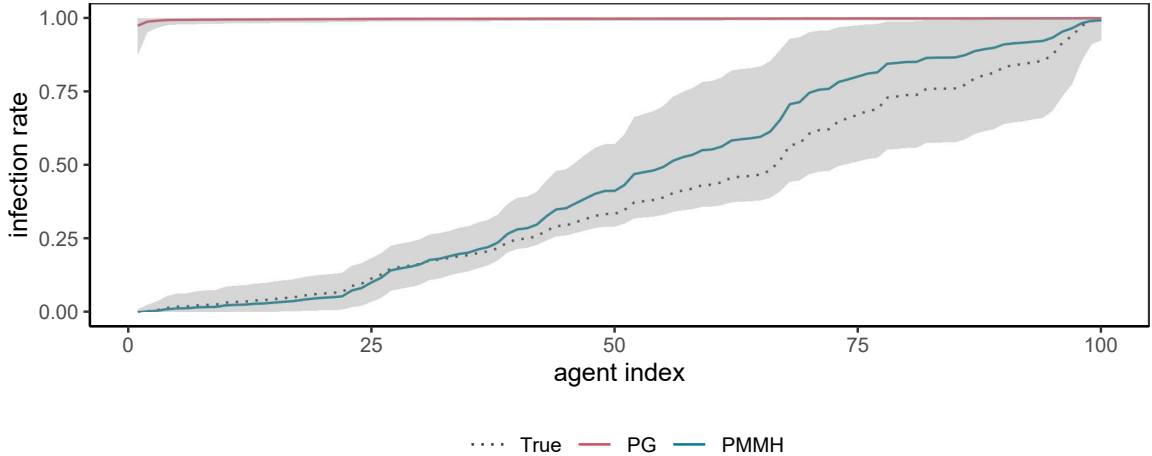}
        \caption{Simulation study 3: The estimated individual infection rates from PG (red) and PMMH (blue) with 95\% CI. A gray dotted line represents the true infection rates}\label{fig:PG}.  
\end{figure} 

\subsection{Simulation 3 - Comparison between PG and PMMH}\label{sec:sim_PG_PMMH}

In this section, we compare the performance of PG and PMMH algorithm since PG may be preferred when conjugate priors are available and \cite{Lindsten2012} show that the PG outperforms PMMH using the same proposal density of parameters. For the PG implementation, conjugate Beta prior is assigned to reporting rate $\rho$ and remaining parameters are sampled in a Metropolis-Hastings fashion, in accordance with Metropolis within particle Gibbs \citep{Lindsten2012}. Under $N=100$, $T=100$ and $P=100$ setting, the knowledge-based priors, which is described in Sub-section \ref{sec:prior}, are assigned to parameters and $\operatorname{Beta}(1,1)$ is used for $\rho$ in PG implementation. The estimated infection rates are illustrated in Figure \ref{fig:PG} and posterior means of all parameters with 95\% CI are represented in Appendix (Table \ref{table:PG}). 

The PG algorithm, which samples the parameters conditioned on the states and vice versa, fails to recover the parameters resulting in the nearly identical infection rates for all agents. This is because there exists a strong dependence between parameters and hidden states in our agent-based SIS model and so a poor mixing is produced by alternating between updates of parameters and hidden states.

\subsection{Simulation 4 - Both infection and recovery rate are unknown}\label{sec:sim_fully}

In the final sets of simulations, we consider both infection and recovery rate are unknown such that we are now interested in estimating $\theta =(\beta_{\alpha_0},\beta_{\alpha_1},\beta_{\lambda_0},\beta_{\lambda_1},\beta_{\gamma_0},\beta_{\gamma_1},\rho)$. Under $N=500$, $T=30$ and $P=100$, the recovery rate is generated with $\beta_{\gamma}=(-1, -1)$.  The truncated Normal distributions are assigned to $\beta_{\lambda_1}$ and $\beta_{\gamma_1}$; $\beta_{\lambda_1}\sim N_+(0,3^2)$ and $\beta_{\gamma_1}\sim N_-(0,3^2)$. 

\begin{figure}[t]
    \centering
        \includegraphics[width = 1 \linewidth]{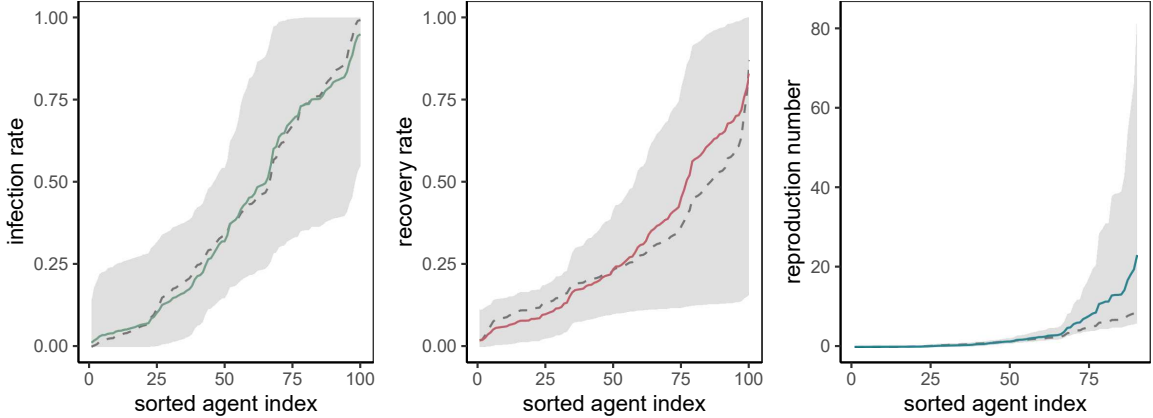}
        \caption{Simulation study 4: Posterior estimates of infection rates and recovery rates for each agent (left) and the estimated reproduction number $R^n = \lambda^n/\gamma^n$(right). The data generating values are represented with a gray dotted line.}\label{fig:RR}
\end{figure}

Figure \ref{fig:RR} illustrates the estimated infection rate, recovery rates and reproduction number $R^n = \lambda^n/\gamma^n$ with 95\% CI for each agent. The reproduction number is truncated to 90 agents since large posterior uncertainty from remaining 10 agents hinders visualization. The reproduction number for all agents is shown in Appendix (Figure \ref{fig:reproductive_all}). 

The posterior means of infection and recovery rate are close to the data generating values which is unique to each agent. While there is large posterior uncertainty for agents with high recovery rates, the range of CI of infection rate remain similar over time and the posterior mean of recovery and infection rates is close to the true data generating values.

\subsection{Summary of simulations}
%Based on the simulation described above, we summarize the main findings in this section. 
We conduct four sets of simulation studies to explore operating characteristics of the HMM approach for agent-based SIS model. PMMH performance is examined in relation to particle number, time points, number of agents, and prior distribution selection. Furthermore, we compare the performance of PG and PMMH, and extend the model to the case when both infection and recovery rates are unknown. Our findings from the four simulation studies are summarized below.

In our first set of simulations, issues relating to lack of identifiability, multimodality, and large posterior uncertainty are addressed when using an uninformative Normal prior distribution. The uninformative prior can lead to a diffuse prior of hidden states resulting in the particles spreading out too much and poor mixing in MCMC chain except for a large number of time points \citep{Wigren2019}. In spite of these issues leading to poor parameter estimations, estimated (aggregated) counts are still close to observations under all simulation settings. Therefore evaluation based only on aggregated counts is not recommended.

Second, the use of truncated Normal priors, which restrict the sign of parameters based on the knowledge or references and are still relatively uninformative, improve parameter estimation significantly, resolving identifiability and multimodality issues. After isolating these issues, we find that increasing the number of time points or agents results in posterior means closer to the true values with less posterior uncertainty whereas the number of particles does not have a significant impact on parameter estimation. Also, agent-based SIS model performs well for estimating group-specific infection rates, where the covariate $z^n$ is defined as a binary variable. 

With respect to the comparison between PMMH and PG, the PG produces a poor mixing produced by alternating between updates of parameters and hidden states. This result reconfirms that PMMH outperforms PG under strong dependence between parameters and hidden states \citep{Lindsten2011}. Furthermore, when the model is extended to the case when recovery rate is also unknown, 95\% CI of infection and recovery rates includes the true data generating values which is unique to each agent.

\section{Analysis of COVID-19 Spread in Diamond Princess cruise}\label{sec:real}
%An outbreak of COVID-19 on the Diamond Princess (DP) cruise ship has provided empirical data for the study of its original transmissibility in isolated environments. 
On February 1, 2020, an outbreak of COVID-19 was reported on Diamond Princess cruise ship off the Japanese coast, with a confirmed case identified in an 80-year-old male passenger \citep{wiki,Moriarty2020}. The first case had embarked and became symptomatic on January 20, and the following 10 cases were reported on February 5. While there was an attempt to test all passengers starting on February 15, Covid-19 testing was initially limited to people with symptoms extending to high-risk individuals, such as elderly passengers and people with chronic illnesses. On February 20, the decision was made to evacuate and more than 3,000 passengers left the ship \citep{Nakazawa2020}. 
It was stated that, of the 3711 people on board, 712 (567 passengers and 145 crew) reported confirmed infection with the virus and, including 381 symptomatic cases at the time of testing by late March \citep{Moriarty2020}. The confirmed cases by age group on February 20 are reported and, of the 619 confirmed cases, 154 and 465 cases are observed for younger people ($<$60 years old) and elderly people ($\geq$ 60 years old) respectively \citep{covidData1,covidData2,rocklov2020}.

%The 3711 people (2666 passengers and 1045 crew members) were quarantined in their cabins while the crew continued to work\citep{Kakimoto2020}.

\begin{figure}[t]
    \centering
        \includegraphics[width=1 \linewidth]{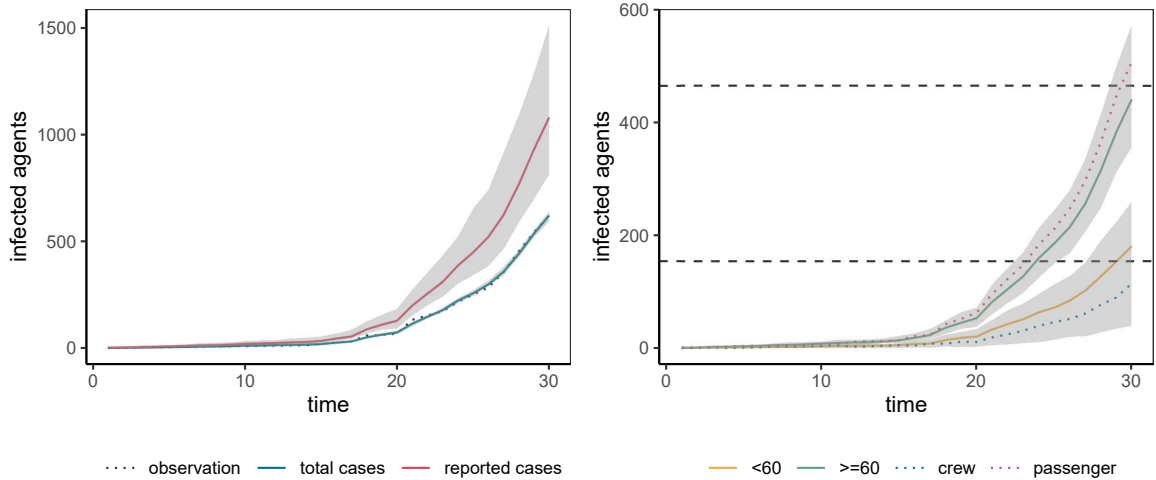}
\caption{Diamond princess data analysis: The estimated confirmed cases for 30 days on Diamond Princess cruise ship. In the left panel, the blue solid line and dotted line represent the estimated reported cases and the observed cases respectively. The red solid line represents the estimated actual cases when all are assumed to be tested. In the left panel, the estimated confirmed cases by age group are shown with different color with 95\% CI and The dotted lines indicates the estimated cases by crew and passengers over time. The reported confirmed cases by age group on the day 30 are depicted by the horizontal lines.}\label{fig:real}
\end{figure}

Since the outbreak of COVID-19 on the Diamond Princess occurred in a closed environments, it allows us to understand the association of demographic characteristics and the transmission of novel COVID-19 prior to implementation of any external interventions using proposed agent-based SIS model. Thus, we examine age-specific infectiousness by comparing trends in infections among elderly ($\geq$60 years) versus younger ($<$60 years) passengers. We further consider two different networks of connections in the cruise ship; one among the passengers and the other among the crew, as they used different levels of deck in the isolated setting and the studies on different transmission rates between passenger and crew are conducted \citep{Lai2021,Jennes2021,rocklov2020}. We therefore examine different infection rates between passengers and crew. 

January 21 was set as day 1, as January 20 was the day of departure (day 0), and February 19 (day 30) was set as the last day as most passengers disembarked on this day. We generate 3711 agents with 1546 younger and 2165 elderly agents indicated by a binary variable $z^n$. Among 3711 agents, 1045 (2666) agents are assigned to crew (passengers) to consider two separate networks. That is, the crew neighborhood $\mathcal{N}_{crew}$ with $\mathcal{D}(n)=1045$ and $\mathcal{N}_{passengers}$ with $\mathcal{D}_{crew}(n)=2666$ are defined. All attributes of agents are consistent with the reported demographic information on the Diamond Princess cruise ship \citep{wiki}.

We use the confirmed cumulative cases as the response variable $y_t$ for days $t=1,\ldots,30$. Following the suggestion from \cite{Lai2021}, linear interpolation was used to compute observed infection cases for dates when the number of confirmed cases were not available. To implement PMMH, we use relatively uninformative, diffuse Normal prior distributions with standard deviation of 3 on  ($\beta_{\alpha0},\beta_{\alpha1},\beta_{\lambda0}$). The truncated Normal prior of $\mathcal{N}_+(0,3^2)$ is assigned to $\beta_{\lambda1}$ as elderly individuals are more likely to be infected according to preliminary studies \citep{Davies2020,KOBAYASHI20221}. Also, the transformed parameter logit($\rho$) is given a Normal prior distribution with mean logit(0.8) and standard deviation of 1 to make $\rho$ falls between 0.36 and 0.97 with probability 0.95. Since previous studies suggest the average time until recovery is $13.5$ days \citep{Ling2020} approximately, which is interpreted as $1/\lambda$, we fix the recovery rate with 1/13.5. We use P = 100 particles and a Normal random walk proposal transition is implemented to jointly update the parameters ($\beta_{\alpha0},\beta_{\alpha1},\beta_{\lambda0}, \beta_{\lambda1}, logit(\rho)$), with a standard deviation of 0.1 to achieve suitable acceptance probabilities. We fit the agent-based SIS model with 10000 MCMC iterations after a burn-in 10000 iterations.

\begin{table}[t]
\centering
\begin{tabular}{rrrrrr}
  \hline
 & $\beta_{\alpha_0}$ & $\beta_{\alpha_1}$ & $\beta_{\lambda_0}$ & $\beta_{\lambda_1}$ & $\rho$ \\ 
  \hline
mean & -7.46 & -0.17 & -1.32 & 1.09 & 0.59 \\ 
  2.5\% & -9.98 & -3.09 & -3.11 & 0.02 & 0.41 \\ 
  97.5\% & -6.22 & 2.95 & -0.53 & 3.51 & 0.76 \\ 
   \hline
\end{tabular}
\caption{Diamond princess data analysis: The posterior estimates of parameters from Diamond Princess data with 95\% CI. }\label{table:real}
\end{table}

The left panel of Figure \ref{fig:real} represents estimated reported cases with 95\% CI over time. Also, by the consideration of reporting rate $\rho$, it is possible to estimate all confirmed cases in addition to reported cases when all people aboard are assumed to be tested. There is a large posterior uncertainty for all confirmed cases as time progresses while the estimated reported cases shows narrow CI for all 30 days. %This is because PF is designed to identify combinations between the reporting rate and all confirmed cases that match the observation $y_t$. In the case of overestimation of all confirmed cases, the reporting rate is underestimated to be in accordance with observation, and vice versa. 
The right panel of Figure \ref{fig:real} shows estimated reported cases by age group with 95\% CI. The predicted cumulative number of infected elderly people and younger people on February 19 is 437 out of 2165 (20.18\%) and 183 out of 1546 (11.83\%) respectively and the corresponding observed cases, shown as vertical lines, fall into the 95\% CI. Additionally, The predicted cumulative number of crew and passengers are illustrated with dotted lines as underlying contact networks are structured according to crew-crew and passenger-passenger. In accordance with observations, the majority of infections occur among passengers. Furthermore, based on the posterior estimates (Table \ref{table:real}), the reproduction number can be computed for each age group, and interpreted as i) The expected number of cases directly generated by one younger infected person is 3.09 (95\% CI: (0.58, 5.00)) ii) the expected number of cases directly generated by a elderly infected person is 5.99 (95\% CI: (4.50, 8.05)). The posterior estimates of $\rho$  indicates that 40.8\% of the total cases are not reported as a whole in the Diamond Princess data.

\section{Discussion}
In this paper, we developed a HMM approach to provide an approximate inference for agent-based SIS model designed for identifying underlying dynamics when observation at individual level is not available. Further, we illustrated how to estimate hidden agent states and the parameters using PMCMC algorithm and examined the performance of the agent-based SIS model under a variety of data generation and prior assumption settings. Our results indicate that there is a strong dependency between updates of parameters and hidden states, and knowledge-based priors are essential to resolve the inferential challenges arising from ABM rather than large numbers of agents and particles. 

Next, we applied the proposed approach to describe COVID-19 infection dynamics overall and by age group in the Diamond Princess cruise ship. In Diamond Princess data analysis, the prediction from agent-based SIS model successfully captures the observed aggregate counts and the age-specific infectiousness. While several studies have investigated the application of the compartment models or examined heterogeneous transmission of COVID-19 in the Diamond Princess cruise \citep{Lai2021,KOBAYASHI20221,rocklov2020}, we described the dynamic process through simulations of interaction between individuals according to the set of rules and estimated the underlying parameters with uncertainty quantification. %Therefore, our approach can be effective in forecasting a new epidemic outbreak for which no previous studies have been conducted.

%it is not always possible to find optimal parameters from the literature, especially for emergent phenomena for which no previous studies have been conducted.
%It is worth noting that it is challenging to impute parameter values for simulating the behavior of agents in an ABM, For example, infection rates are often derived from other related studies or set for scenarios using corresponding parameter values. However, it is not always possible to find optimal parameters from the literature, especially for emergent phenomena for which no previous studies have been conducted. Our aim is to examine and improve parameter inference based on observed data with HMM approach for ABM, and to quantify the associated uncertainty in estimation. 

%In our paper, we focused on parameter inference using observed data for ABM with HMM approach and assessed sensitivity to assumptions on prior distribution.
It is worth noting that one of the most challenging steps in building an ABM is identifying parameter values for simulating the behavior of agents, which are commonly derived from other relevant studies. It is not always possible, however, to find optimal parameters based on existing literature, especially for emerging phenomena that have not previously been studied. Therefore, our contributions are to alleviate inferential challenges for ABM which can describe natural world in a flexible manner and to provide quantification of the uncertainty in parameter values within a Bayesian framework.

For future work, more complex compartmental models (e.g susceptible-exposed-infected-recovered or dead) and complex networks can be considered to accommodate more realistic dynamic process of disease transmission. While we attempted to explore the operating characteristics of the PMCMC approach under the simplified rules for agent behaviors, a larger parameter space can be explored by choosing optimal prior or proposal distributions.

\subsection*{Acknowledgement}
We thank Dr. Anna Bershteyn for providing valuable comments. Drs. Adhikari and Um were supported by funding from Johnson \& Johnson Women in Stem Award to Dr. Adhikari.

%%%%%%%%%%%%%%%%%%%%%%%%%%%%%%%%%%%%%%  Reference
%\nocite{*} 

\bibliographystyle{abbrvnat}
\bibliography{Arxiv}

\appendix

\begin{algorithm}[h]
\caption{Particle Gibbs}
%\textbf{Input:} Trajectory $x_{1: T}^{\prime}[r]$ and parameter $\mathbf{\theta} \in \Theta$.\\
%\textbf{Output:} Trajectory $x_{1: T}^{\prime}[r+1]$
\begin{algorithmic}[1] 
    \State Set the initial value $\theta[0]$ and $\bx_{1: T}^\prime[0]$ arbitrarily
\For{$m=1$ to $M$}
\State Set $\theta = \theta[m]$
\State Draw $\bx_{0}^{(p)} \sim p_{\theta}\left(\bX_{0}\right)$ for $p=1, \cdots, P-1$.
    \State Set $\bx_{0}^{(P)}=\bx_{0}^{\prime}[m-1]$
    \State Calculate weight $w_{1}\left(\bx_0^{(p)}\right)=p_{\theta}\left(\by_{0} \mid \bx_{0}^{(p)}\right)$ for $p=1, \cdots, P$
    \State Compute normalized weight $\bar{w}_{0}^{(p)}=w_{0}^{(p)} / \sum_{j=1}^{P} w_{0}^{(j)}$ for $p = 1,\cdots,P$
    \For{$t=1$ to $T$}
        \State Draw $a_{t-1}^{(p)}\sim \mathcal{C}(\{\bar{w}^{(p)}_{t-1}\}_{p=1}^{P})$ for $p=1,\cdots,P-1$ where $\mathcal{C}$ is the categorical distribution.
            %     \If{Ancestor sampling == TRUE} 
            %     \State Draw $a_t^P$ with $\mathbb{P}(a_t^P=j)\propto w_{t-1}^{j} p_{\theta}\left(x_{t}^{\prime} \mid x_{t-1}^{j}\right)$
            % \Else
            %     \State Set $a_{t}^{P}=P$
            % \EndIf
        %\State Set $a_{t}^{(P)}=P$
        \State Draw $\bx_t^{(p)}\sim p(\bx_t | \bx_{t-1}^{a_{t-1}^{(p)}},\by_t)$ for $p=1,\cdots,P-1$.
        
        \State Set $a_{t=1}^{(P)}=P$ and $\bx_t^{(P)} = \bx_t^\prime[m-1]$
        \State Set $\bx_{0:T}^{(p)} = \{\bx_{1:t-1}^{a_t^{(p)}},\bx_t^{(p)}\}$ for $p=1,\cdots, P$
        \State Calculate weight $w_{t}^{(p)}\left(\bx_{t}^{(p)}\right) = p_{\theta}\left(\by_{t} \mid \bx_{t}^{(p)}\right) $ for $p=1, \cdots, P$
        \State Compute normalized weight $\bar{w}_{t}^{(p)}=w_{t}\left(\bx_{t}^{(p)}\right) / \sum_{j=1}^{P} w_{t}\left(\bx_{t}^{(p)}\right)$ for $p=1, \cdots, P$
    \EndFor
    \textbf{end for}
    \State Sample $k$ with $P(k=p) = \bar{w}_{T}^{(p)}$.
    \State Set the reference trajectory $\bx_{1: T}^{\prime}[m] = \bx_{1: T}^{k}$
    \State Draw $\theta[m] \sim p\left(\theta \mid \bx_{1: T}^\prime[m], \by_{1: T}\right)$
\EndFor\textbf{end for}
\end{algorithmic}\label{algo:cSMC}
\end{algorithm}

\begin{figure}[t]
    \centering
        \includegraphics[width=1 \linewidth]{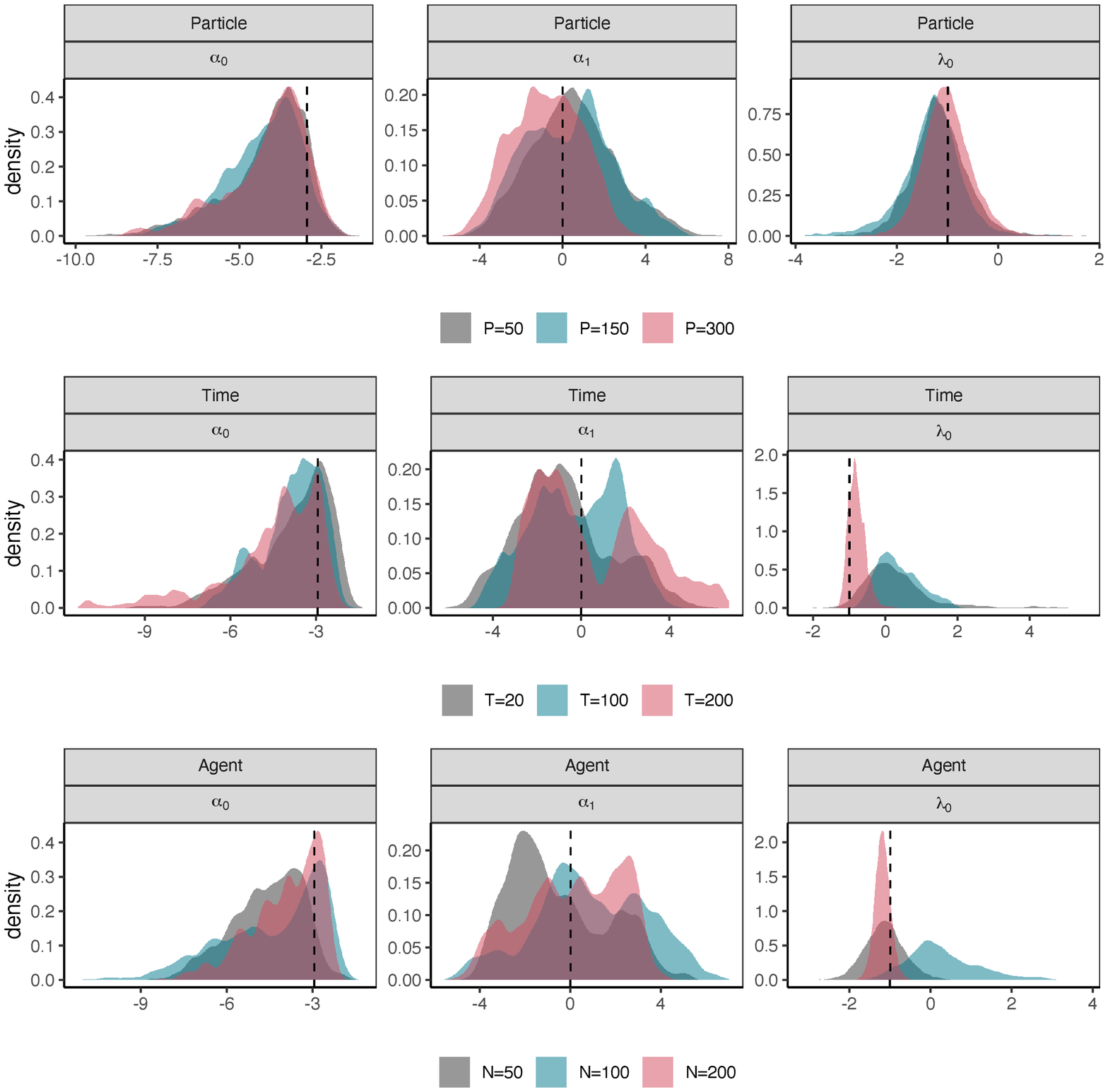}
    \caption{Simulation study 1 : The posterior distributions with Normal priors under 9 different settings. i) the number of particles varies $P\in(50, 150,300)$ at N=50 and T=30 ii) the number of time points varies $T\in(20,100,200)$ at N=100 and P=100 iii) the number of agents varies $N\in(50,100,200)$ at T=30 and P=100.}\label{fig:remain_uninfo}
\end{figure}

\begin{table}[h]
\centering
\begin{tabular}{lllllll}
  \toprule
 & & $\beta_{\alpha0}$ & $\beta_{\alpha1}$ & $\beta_{\lambda0}$ & $\beta_{\lambda1}$ & $\rho$ \\ 
  \toprule
 (N,T,P)   & True & -2.99 & 0 & -1 & 2 & 0.8\\ 
   \midrule
(50,30,50) &mean & -4.19 & 0.52 & -1.17 & -0.17 & 0.73\\
  &95\%CI & (-7.37,-2.47) & (-3.24,4.86) & (-2.24,-0.00) & (-6.58,6.99)  & (0.47,0.91) \\ \rule{0pt}{3ex}   
(50,30,150)&  mean & -4.30 & 0.32 & -1.32 & -1.20 & 0.74 \\ 
  & 95\%CI & (-7.06,-2.44) & (-3.35,4.50) & (-2.70,-0.19) & (-8.63,4.35) & (0.49,0.92) \\ \rule{0pt}{3ex}   
 (50,30,300)  &  mean & -4.18 & -0.80 & -1.01 & 1.64 & 0.71 \\ 
  &95\%CI & (-7.23,-2.40) & (-3.96,2.24) & (-1.91,0.01) & (-3.54,5.56) & (0.45,0.90) \\ 
 \midrule
 (50,20,100) &  mean & -3.86 & -0.76 & 0.30 & -1.09 & 0.64 \\ 
  &95\%CI & (-7.13,-2.06) & (-4.54,3.74) & (-1.00,2.87) & (-6.89,4.58) & (0.47,0.85) \\ \rule{0pt}{3ex}  
 (50,100,100) &  mean & -3.85 & -0.27 & 0.36 & -2.85 & 0.74 \\ 
  & 95\%CI & (-6.17,-2.31) & (-4.01,3.05) & (-0.57,1.68) & (-4.89,-1.20) & (0.66,0.82) \\ \rule{0pt}{3ex} 
 (50,200,100) &  mean & -4.55 & 0.56 & -0.80 & 2.79 & 0.78 \\ 
  & 95\%CI & (-9.24,-2.49) & (-2.84,5.92) & (-1.17,-0.28) & (1.25,5.01) & (0.70,0.84) \\ 
 \midrule
(50,30,100)&  mean & -4.59 & -0.54 & -1.14 & 0.32 & 0.72 \\ 
  &95\%CI & (-7.29,-2.62) & (-3.69,3.72) & (-2.11,-0.08) & (-5.64,6.36) & (0.47,0.91) \\ \rule{0pt}{3ex}  
(100,30,100) &   mean & -4.43 & 0.89 & 0.29 & 2.21 & 0.59 \\ 
  &95\%CI & (-8.44,-2.15) & (-4.19,5.31) & (-1.17,2.47) & (-4.21,6.00) & (0.46,0.77) \\ \rule{0pt}{3ex}  
(200,30,100) &  mean & -3.95 & 0.10 & -1.19 & -0.56 & 0.83 \\ 
  &95\%CI & (-6.86,-2.35) & (-3.99,3.32) & (-1.57,-0.79) & (-3.73,3.23) & (0.64,0.95) \\
  \bottomrule
\end{tabular}
\caption{Simulation study 1 : The posterior mean of parameters with 95\% CI compared to the data generating values when Normal priors are used. The 9 different simulation setting are examined; $P\in(50,150,300)$ with $N=100$ and $T=30$, $T\in(30,100,200)$ with $P=100$ and $N=100$ and $N\in (50,100,200)$ with $T=30$ and $P=100$.}\label{table:comprehensive_Normal}
\end{table}

\begin{figure}[t]
    \centering
        \includegraphics[width=1 \linewidth]{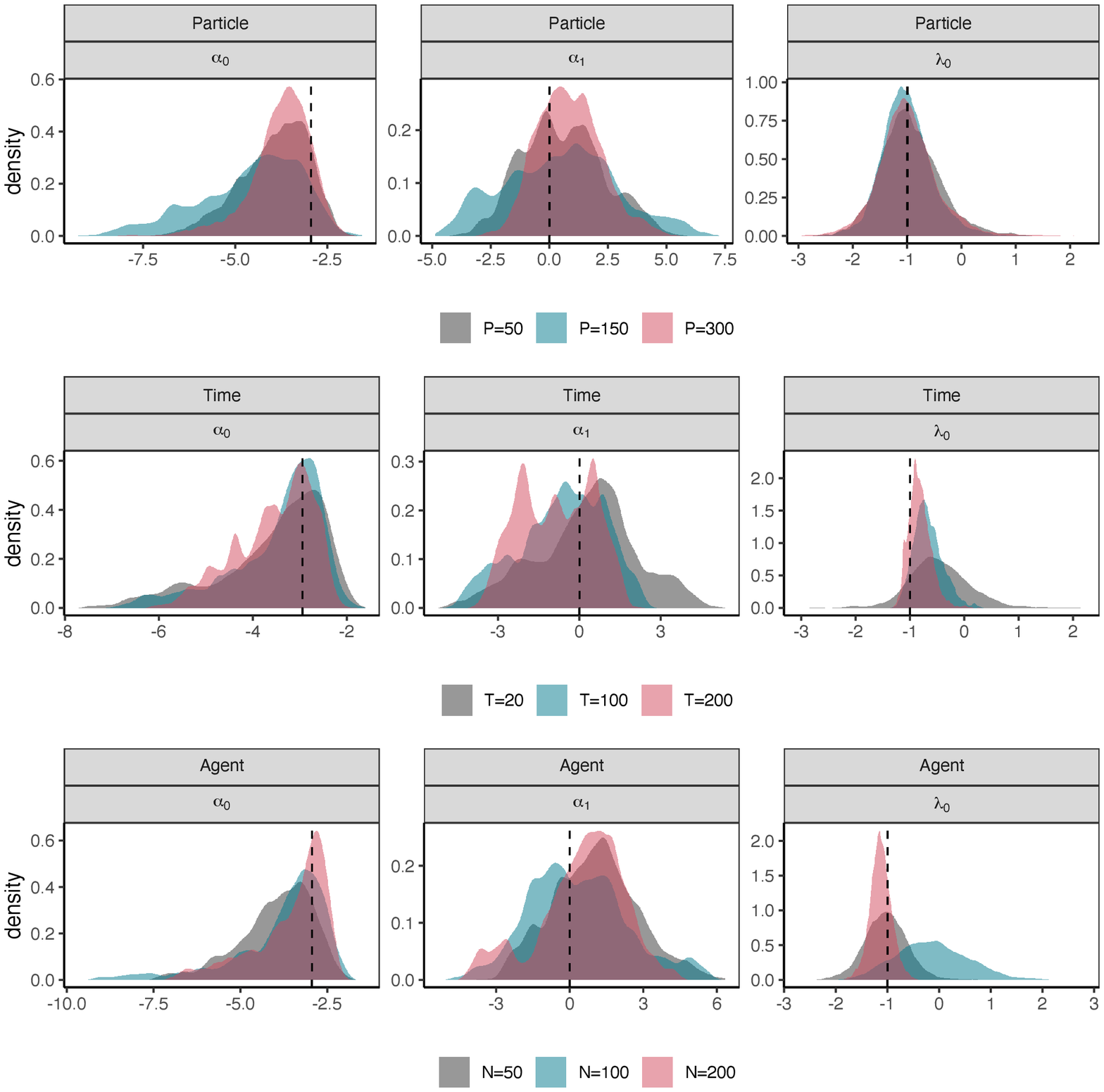}
    \caption{Simulation study 2 : The posterior distributions with knowledge-based priors under 9 different settings. i) the number of particles varies $P\in(50, 150,300)$ at N=50 and T=30 ii) the number of time points varies $T\in(20,100,200)$ at N=100 and P=100 iii) the number of agents varies $N\in(50,100,200)$ at T=30 and P=100.}\label{fig:remain_knowledge}
\end{figure}

\begin{table}[h]
\centering
\begin{tabular}{lllllll}
  \toprule
(N,T,P) & & $\beta_{\alpha0}$ & $\beta_{\alpha1}$ & $\beta_{\lambda0}$ & $\beta_{\lambda1}$ & $\rho$ \\ 
  \toprule
   & True & -2.99 & 0 & -1 & 2 & 0.8\\ 
   \midrule
 (50,30,50)  & mean & -3.94 & 0.57 & -0.95 & 4.30 & 0.74 \\ 
  & 95\%CI & (-6.04,-2.45) & (-2.55,4.12)  & (-1.93,0.26) & (0.30,9.07) & (0.53,0.88) \\\rule{0pt}{3ex}   
(50,30,150) & mean & -4.61 & 0.48 & -1.02 & 3.41 & 0.75 \\ 
  & 95\%CI & (-7.83,-2.52) & (-3.75,5.41) & (-1.86,-0.00) & (0.45,8.15) & (0.56,0.90) \\\rule{0pt}{3ex}   
 (50,30,300) & mean & -3.76 & 0.83 & -0.99 & 3.55 & 0.75 \\  
 & 95\%CI & (-5.86,-2.49) & (-1.54,3.81) & (-2.02,0.22) & (0.29,8.43) & (0.55,0.90) \\ 
 \midrule
 (50,20,100)  &  mean & -3.58 & 0.28 & -0.46 & 3.03 & 0.67 \\ 
 & 95\%CI & (-6.44,-2.12) & (-3.61,3.80) & (-1.58,0.76) & (0.19,7.81) & (0.49,0.84) \\\rule{0pt}{3ex}  
(50,100,100) & mean & -3.49 & -0.66 & -0.67 & 2.37 & 0.74 \\ 
 & 95\%CI & (-6.17,-2.24) & (-3.93,1.98) & (-1.12,-0.07) & (0.75,4.81) & (0.66,0.83) \\\rule{0pt}{3ex} 
 (50,200,100)  &  mean & -3.55 & -0.86 & -0.82 & 2.56 & 0.78 \\ 
 & 95\%CI& (-5.37,-2.40) & (-3.18,1.39) & (-1.15,-0.36) & (1.12,4.36) & (0.71,0.84) \\ 
 \midrule
(50,30,100) & mean & -3.98 & 1.02 & -1.03 & 2.97 & 0.74 \\ 
 &95\%CI & (-6.61,-2.36) & (-2.29,4.74) & (-1.88,-0.13) & (0.23,6.91) & (0.52,0.90) \\\rule{0pt}{3ex}  
(100,30,100)  &  mean & -3.83 & 0.34 & -0.13 & 3.23 & 0.63 \\  
 & 95\%CI & (-7.93,-2.20) & -(3.34,4.89) & (-1.42,1.39)  & (0.52,6.19) & (0.47,0.80) \\\rule{0pt}{3ex}  
(200,30,100)  & mean & -3.56 & 0.50 & -1.14 & 1.78 & 0.81 \\ 
 & 95\%CI & (-6.37,-2.30) & (-3.60,3.44) & (-1.57,-0.70) & (0.17,3.97) & (0.64,0.94) \\  
  \bottomrule
\end{tabular}
\caption{Simulation study 2 : The posterior mean of parameters with 95\% CI compared to the data generating values when knowledge-based priors are used. The 9 different simulation setting are examined; $P\in(50,150,300)$ with $N=100$ and $T=30$, $T\in(30,100,200)$ with $P=100$ and $N=100$ and $N\in (50,100,200)$ with $T=30$ and $P=100$.}\label{table:comprehensive_truncated}
\end{table}

\begin{table}[b]
\centering
{\tabcolsep=3pt
\begin{tabular}{lllllll}
  \toprule
 & $\beta_{\alpha_0}$ & $\beta_{\alpha_1}$ & $\beta_{\lambda_0}$ & $\beta_{\lambda_1}$ & $\rho$ \\ 
  \midrule
 True & -2.99 & 0 & -1 & 2 & 0.8\\ 
 \midrule
 PG & & & & & & \\
 ~~ mean & -2.49 & -0.06 & 6.65 & 0.76 & 0.50 \\ 
  ~~95\% CI & (-4.13, -1.43) & (-2.01, 1.75) & (4.53, 10.64) & (0.03, 2.15) & (0.49, 0.51) \\ 
  \midrule
   PMMH & & & & & & \\
  ~~mean & -4.07 & 0.11 & -0.78 & 2.92 & 0.79 \\ 
  ~~95\% CI & (-8.59,-2.41) & (-3.03, 5.22)  & (-1.15,-0.30) & (1.27, 5.02) & (0.71, 0.84) \\
  \bottomrule
\end{tabular}}
\caption{Simulation study 3 : The posterior estimates with 95\% CI from particle Gibbs (top) and PMMH (bottom)}\label{table:PG}
\end{table}

\begin{figure}[t]
    \centering
        \includegraphics[width=1 \linewidth]{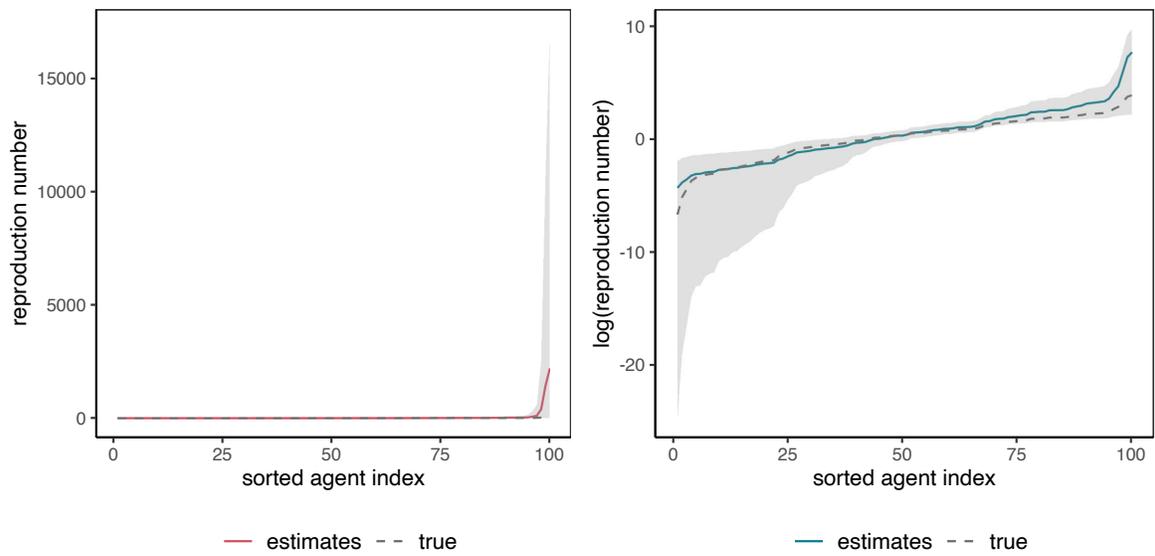}
    \caption{Simulation sutdy 4 : Posterior estimates of reproduction number $R^n = \lambda^n/\gamma^n$ of each agent (left) with 95\% CI and log scales of the estimates (right). The data generating values are represented with a gray dotted line.}\label{fig:reproductive_all}
\end{figure} 

\end{document}